\documentclass[
]{ceurart}
\usepackage{subfig}
\begin{document}

\copyrightyear{2020}
\copyrightclause{Copyright for this paper by its authors.
  Use permitted under Creative Commons License Attribution 4.0
  International (CC BY 4.0).}

\conference{Lloret'21: Indoor Positioning and Indoor Navigation, 29 Nov-2 Dec 2021, Lloret de Mar, Spain}

\title{5G Deployment Strategies for High Positioning Accuracy in Indoor Environments}

\author[1]{{Maria Posluk}}[]
\ead{mariaposluk@gmail.com}
\address[1]{Department of Electrical Engineering, Linköping University, Linköping, Sweden}
\author[1]{{Jesper Ahlander}}[]
\ead{jesper.ahlander@gmail.com}
\author[2]{{Deep Shrestha}}[]
\address[2]{Ericsson Research, Linköping, Sweden}
\ead{deep.shrestha@ericsson.com}
\author[2]{{Sara Modarres Razavi}}[]
\ead{sara.modarres.razavi@ericsson.com}
\author[2]{{Gustav Lindmark}}[]
\ead{gustav.lindmark@ericsson.com}
\author[2]{{Fredrik Gunnarsson}}[]
\ead{fredrik.gunnarsson@ericsson.com}

\begin{abstract}
Indoor positioning is currently recognized as one of
the important features in emergency, commercial and industrial
applications. The 5G network enhances mobility, flexibility,
reliability, and security to new higher levels which greatly
benefit the IoT and industrial applications. Industrial IoT (IIoT)
use-cases are characterized by ambitious system requirements
for positioning accuracy in many verticals. For example, on
the factory floor, it is important to locate assets and moving
objects such as forklifts. The deployment design for different
IIoT environments has a significant impact on the positioning
performance in terms of both accuracy and availability of the
service. Indoor factory (InF) and indoor open office (IOO) are
two available and standardized {Third Generation Partnership Project} (3GPP) scenarios for evaluation of indoor channel models and positioning performance in IIoT use cases. 
This paper aims to evaluate the positioning performance in terms of
accuracy and availability while considering different deployment strategies. Our simulation-based evaluation shows that deployment plays a vital role when it comes to achieving high accuracy positioning performance. It is for example favorable to deploy the 5G Transmission and Reception Points (TRPs) on the walls of the factory halls than deploying them attached to the ceiling.
\end{abstract}

\maketitle

\section{Introduction}
Localization in cellular networks is primarily used to locate a user equipment (UE) in outdoor-only scenarios, often by exploiting global navigation satellite system (GNSS) that can guarantee meter-level accuracy. In the recent years, mainly since the third generation partnership project (3GPP) Release 13 \cite{Henrik}, the focus on indoor positioning has gained a lot of attention mainly due the updated federal communications commission (FCC) requirements concerning emergency services for indoor calls \cite{FCC} and also to address many commercial use-cases that benefit from positioning information. Moreover, the presence of 5G which has the potential to improve the indoor positioning estimations to sub-meter level accuracy \cite{delPeral2017} and provides an opportunity to enable a plethora of applications in manufacturing industry.   

In the 5G positioning study in 3GPP Release 16 \cite{RP-181399}, the objective was to achieve indoor positioning accuracy below 3m, however the supported 5G positioning features have the technology that has potential to support much more precise positioning. The wide bandwidths, beam-based systems, and higher numerology of 5G compared to 4G all enable improved positioning resolution. Moreover, dense and tailored deployments with small cells and large overlaps improve accuracy and, together with beam-based transmissions, provide more spatial variations that can be exploited for radio frequency fingerprinting \cite{5GSmart}.
  
One important aspect in the outcome of any cellular positioning method is the deployment of the nodes which perform the transmission and reception of the signals. It is well-known that the number of nodes and how they are distributed within the serving area has a great impact on the positioning accuracy. Due to this dependency, there is always a trade-off between how accurate the position estimation can be and how complex and costly the positioning deployment is \cite{AccPosUWB}. In 5G architecture the concept of cells are represented by transmission and reception points (TRPs) that transmit reference signals during a positioning occasion to be used by the UEs to perform measurements for UE localization.  When it comes to indoor positioning the TRP deployment becomes even more prominent because the indoor environment contains more obstacles leading to higher non line of sight (NLOS) conditions that add challenges in achieving a high positioning accuracy \cite{PastPresentFuture}.

There is already an extensive research on the topic of indoor positioning exploiting different technologies. However, most of these studies aim to explore and study the accuracy and the potential of each indoor positioning solution assuming a fixed deployment which is typically optimized to provide decent communication services. Analyzing the impact of different deployment strategies on positioning performance therefore remains unexplored.

In this paper, our aim is to analyze different aspects of TRP deployment strategy and understand its impact on 5G indoor positioning performance. The results are then analyzed for the purpose of understanding how positioning accuracy and availability relates to different deployments. This paper serves as a comprehensive summary of a well-studied thesis work \cite{thesis}.

To provide a good overview of the conducted study, this paper is organized in six sections. Section \ref{sec:theory} introduces downlink time difference of arrival (DL-TDOA) method which is the considered positioning technique, as well as the two analysis methods Cramér-Rao lower Bbund (CRLB) and geometric dilution of precision (GDOP). Section \ref{sec:scenarios} provides introduction to the indoor open office (IOO) and indoor factory (InF) scenarios as defined by 3GPP. Moving on, simulation results and performance evaluations are presented in sections \ref{sec:sim_results} and \ref{sec:perf_eval} while Section \ref{sec:conclusions} draws some conclusive remarks.

\section{Methodology}
In this section, we first define the DL-TDOA method and further explain how we assess a lower bound on the positioning accuracy. Finally, we explain how we evaluate the impact of deployment geometry independent and decoupled from other positioning errors.

\subsection{Downlink Time Difference of Arrival (DL-TDOA) based Positioning} \label{sec:theory}
The 5G DL-TDOA is a positioning method, in which the UE measures the downlink time of arrival (TOA) of positioning reference signals (PRSs) from different TRPs, and reports the time difference of arrival measurements for these TRPs to the location server in relation to the timing of a reference TRP. The positioning estimation is then done by performing multilateration based on the UE reported timing measurements. 

Once the timing measurements are obtained, they are translated to distance observations by multiplying it with the speed of light.
A general measurement equation at time $t$ has the form, 
\begin{equation} \label{eq:theory:meas_model}
y_t = h(\theta_t) + e_t, 
\end{equation} 
where $y_t$ is the measurement, $h(\theta_t)$ is a nonlinear measurement model and $e_t$ is noise. All mentioned variables are vectors. The variable $\theta_t = [x_t \; y_t \; z_t]^T$ is the 3D position of the UE. In general, the function $h(\theta_t)$ will implicitly depend on the known positions of the $N$ TRPs, $p^i = [x^i \; y^i \; z^i]^T$, $i \in \{1,2,\dots,N\}$.

A TOA observation based on TRP $i \in \{1,2,\dots,N\}$ in an asynchronous network can be expressed by,
\begin{equation}\label{eq:theory:toa}
	y_{\text{TOA}}^i = |\theta - p^i| + \delta^i + e^i,
\end{equation}
where $\delta^i$ is the unknown clock bias between the UE and TRP $i$ \cite{SigProc}. 
If the TRPs are synchronized with each other but not with the UE, then $\delta^i = \delta^j\ \forall i,j \in \{1,2,\dots,N\}$.
In this case a DL-TDOA measurement is obtained by taking the difference between two TOA measurements,
\begin{equation} \label{eq:theory:tdoa}
y_{\text{DL-TDOA}}^{j,\, i} = y_{\text{TOA}}^j - y_{\text{TOA}}^i = 
|\theta - p^j| - |\theta - p^i| +  e^j - e^i,
\end{equation}
where $i$ denotes reference TRP and $j$ denotes another TRP for which timing measurements are performed by the UE.

\subsection{Cramér-Rao Lower Bound (CRLB)}
It is often useful to state a lower bound on the performance of any unbiased estimator. The CRLB which expresses a lower bound on the variance of unbiased estimators of a deterministic (fixed, though unknown) parameter, stating that the variance of any such estimator is at least as high as the inverse of the Fisher information, serves that purpose. In order to compute the CRLB one needs to first determine the {Fisher information matrix} (FIM), $\mathcal{I}(\theta)$. When the measurement noise is {additive white Gaussian noise} (AWGN), the FIM becomes \cite{SigProc},
\begin{equation} \label{eq:theory:fim}
\mathcal{I}(\theta) = H^T(\theta)R^{-1}H(\theta),
\end{equation}
where $H(\theta) = \nabla_\theta h(\theta)$ and $R$ is the measurement noise covariance matrix. The information is additive for independent observations \cite{sensorfusion}. CRLB is finally given by
\begin{equation} \label{eq:theory:crlb}
\text{Cov}(\hat{\theta}) \geq \mathcal{I}^{-1}(\theta).
\end{equation}
In positioning studies, plotting the positioning error in meters is a relevant performance metric and it is achieved by calculating the {root mean square error} (RMSE). A lower bound for the RMSE of an estimator is obtained by taking the square root of the trace of CRLB:
\begin{align}
\label{eq:theory:rmse_3d}
\begin{split}
\text{RMSE} &= \sqrt{\text{E}\left[(x - \hat{x})^2 + (y - \hat{y})^2  + (z - \hat{z})^2\right]} \\ &\geq  \sqrt{\text{tr}(\mathcal{I}^{-1}(\theta))}.
\end{split}
\end{align}
When only the horizontal UE positioning error is of interest, then a lower bound can be obtained from the FIM as
\begin{align}
\label{eq:theory:rmse_2d}
\begin{split}
\text{RMSE} &= \sqrt{\text{E}\left[(x - \hat{x})^2 + (y - \hat{y})^2\right]} \\ &\geq  \sqrt{\mathcal{I}^{-1}_{1, \, 1}(\theta) + \mathcal{I}^{-1}_{2, \, 2}(\theta)}
\end{split}
\end{align}
with $\mathcal{I}^{-1}_{1, \, 1}(\theta)$ and $\mathcal{I}^{-1}_{2, \, 2}(\theta)$ being the diagonal elements of $\mathcal{I}^{-1}$ corresponding to the $x$ and $y$ directions \cite{SigProc}.

\subsection{Geometric Dilution of Precision (GDOP)}
We seek to decouple the effect of the TRP deployment geometry on the positioning error from the effects of other measurement errors. For that we use the (GDOP), which intends to state how errors in the measurement will affect the final state estimation \cite{gdop}. GDOP can be computed for each position in the deployment area and it depends on the location of the TRPs and the positioning method which is being used, in our case it is DL-TDOA. For a UE placed at $\theta_t = [x_t\ y_t\ z_t]^T$, the positioning error is essentially the product between GDOP at $\theta_t$ and the measurement error and can be expressed as
\begin{equation*}
\Delta{\textrm{(position estimation error)}} = \textrm{GDOP} \times\Delta{\textrm{(measurement error)}}.    
\end{equation*}
For more details regarding GDOP, see e.g. \cite{lowest_gdop}.

\section{Studied Scenarios} \label{sec:scenarios}

In this study two representative indoor scenarios, IOO and InF, defined in 3GPP are considered \cite{TR38855}. The scenario specific parameters are outlined in the sub-sections to follow.

\begin{table*}
	\caption{Descriptions of InF-SH and InF-DH.}
	\label{table:method:inf_description}
	\centering
	\begin{tabular}{ |p{4cm}|p{4.5cm}|p{4.5cm}| } 
		\hline
		\textbf{Scenario parameter} & \textbf{InF-SH} & \textbf{InF-DH}\\
		\hline
		\hline
		Effective clutter height & \multicolumn{2}{c|}{0--10~m} \\
		\hline
		\multirow{2}{*}{External wall and ceiling type} & \multicolumn{2}{p{7cm}|}{Concrete or metal walls and ceiling with metal coated windows.} \\
		\hline
		\multirow{6}{*}{Clutter type} & Big machines composed of regular metallic surfaces. For example: several mixed production areas with open spaces and storage/commissioning areas.
		& Small to medium metallic machinery and objects with irregular structure. For example: assembly and production lines surrounded by mixed small-sized machines. \\
		\hline
		Typical clutter size & 10~m & 2~m \\
		\hline
		Clutter density & $<40\%$ & $\geq40\%$ \\
		\hline
	\end{tabular}
\end{table*}

\subsection{Indoor Open Office (IOO)}
The IOO scenario is defined as a deployment area of dimensions 120~m $\times$ 50~m $\times$ 3~m designed to capture typical indoor environments such as shopping malls and offices. The TRPs in this type of environment are typically mounted on the ceiling at a height of 3 m \cite{TR38901}.

\subsection{Indoor Factory (InF)}
The InF scenario represents factory halls of varying sizes and clutter densities. The area is 120 m $\times$ 60 m with a ceiling height of 5–25 m where the TRPs can be mounted.

The InF scenario has five different variants. Out of the five, two variants named indoor factory-sparse high (InF-SH) and indoor factory-dense high (InF-DH) are chosen for this study as they depict realistic InF scenarios with sparse and dense clutter densities. The sparse clutter option specifies an industrial factory floor whose $<40\%$ of the area is covered by machines, storage shelves, assembly lines, etc. Likewise the dense clutter option specifies an industrial factory floor with $\geq40\%$ of the area covered by machines, storage shelves, assembly lines, etc. Descriptions of both InF scenarios are presented in Table \ref{table:method:inf_description}.

\subsection{Deployment Strategies}
3GPP has defined standard TRP deployments for both IOO and InF. The proposed deployments are mainly motivated to meet the quality of service (QoS) to support communication requirements of IIoT use cases. In IOO scenario 12 TRPs are deployed in two rows and in InF scenario 18 TRPs are deployed in three rows. In both scenarios an inter site distance (ISD) of 20m is maintained in both x and y axes \cite{TR38901}. Inspired by the 3GPP specifications, in this paper we define and analyze the following three deployment strategies:

\begin{figure}[ht!]
	\centering
	\subfloat[][\label{fig:results:standard}Standard deployment (IOO and InF).]{\includegraphics[clip=true,trim={3.5cm 8.5cm 3.5cm 8.5cm},width=0.50\linewidth]{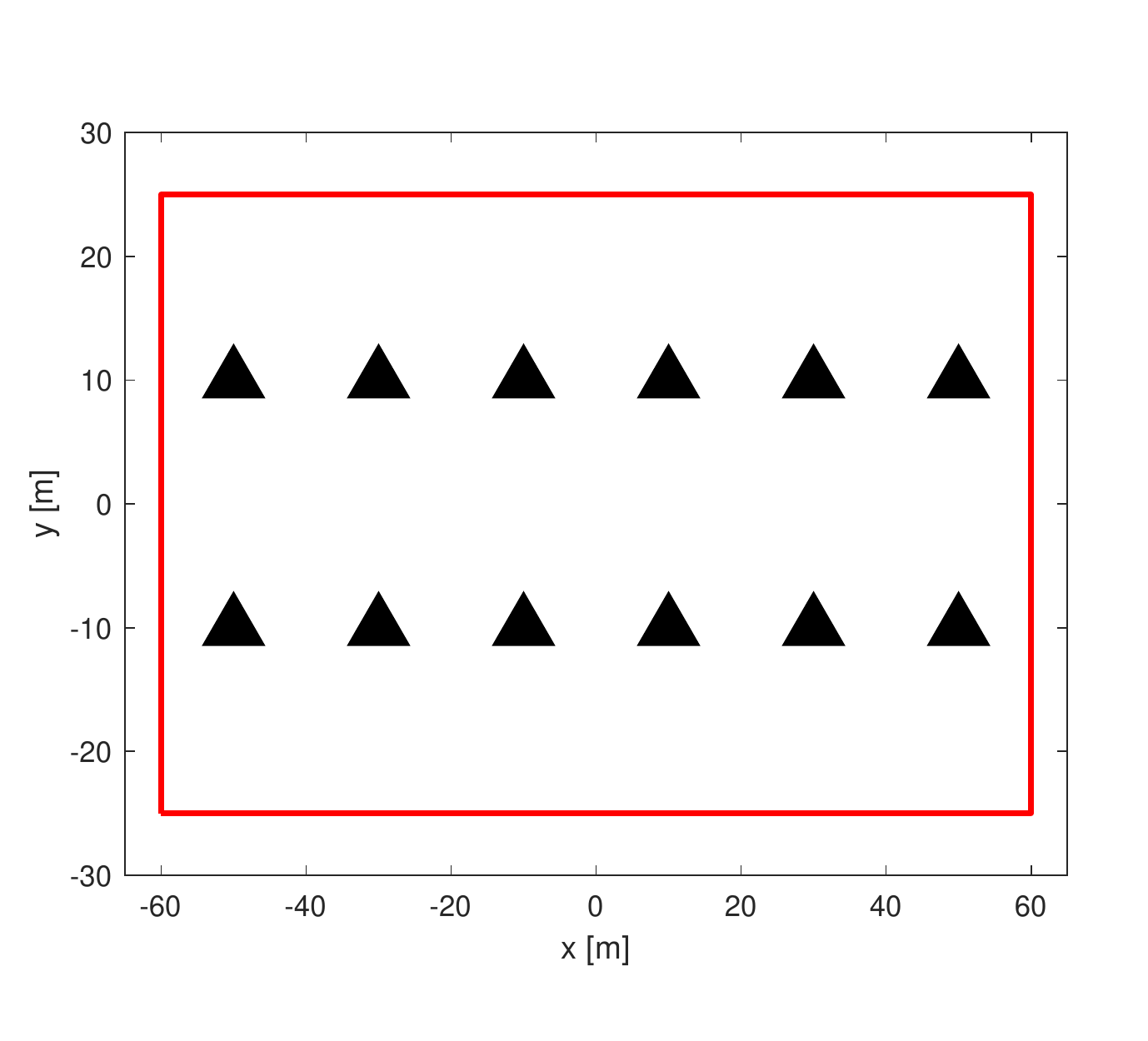}}
	\subfloat[][\label{fig:results:edge}Edge deployment (IOO and InF).]{\includegraphics[clip=true,trim={3.5cm 8.5cm 3.5cm 8.5cm},width=0.50\linewidth]{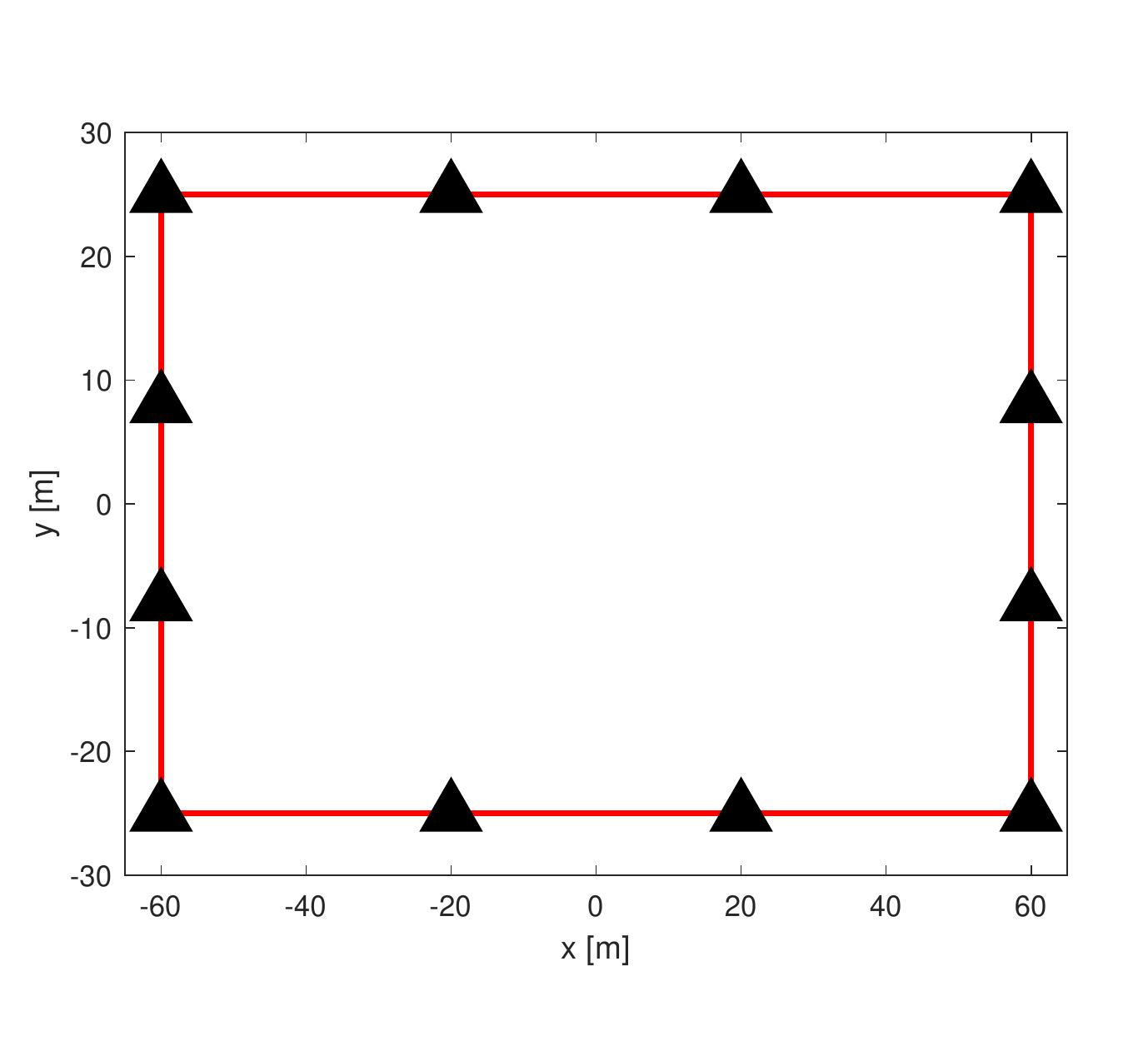}}\\
	\subfloat[][\label{fig:results:mixed_ioo}Mixed deployment for IOO.]{\includegraphics[clip=true,trim={3.5cm 8.5cm 3.5cm 8.5cm},width=0.50\linewidth]{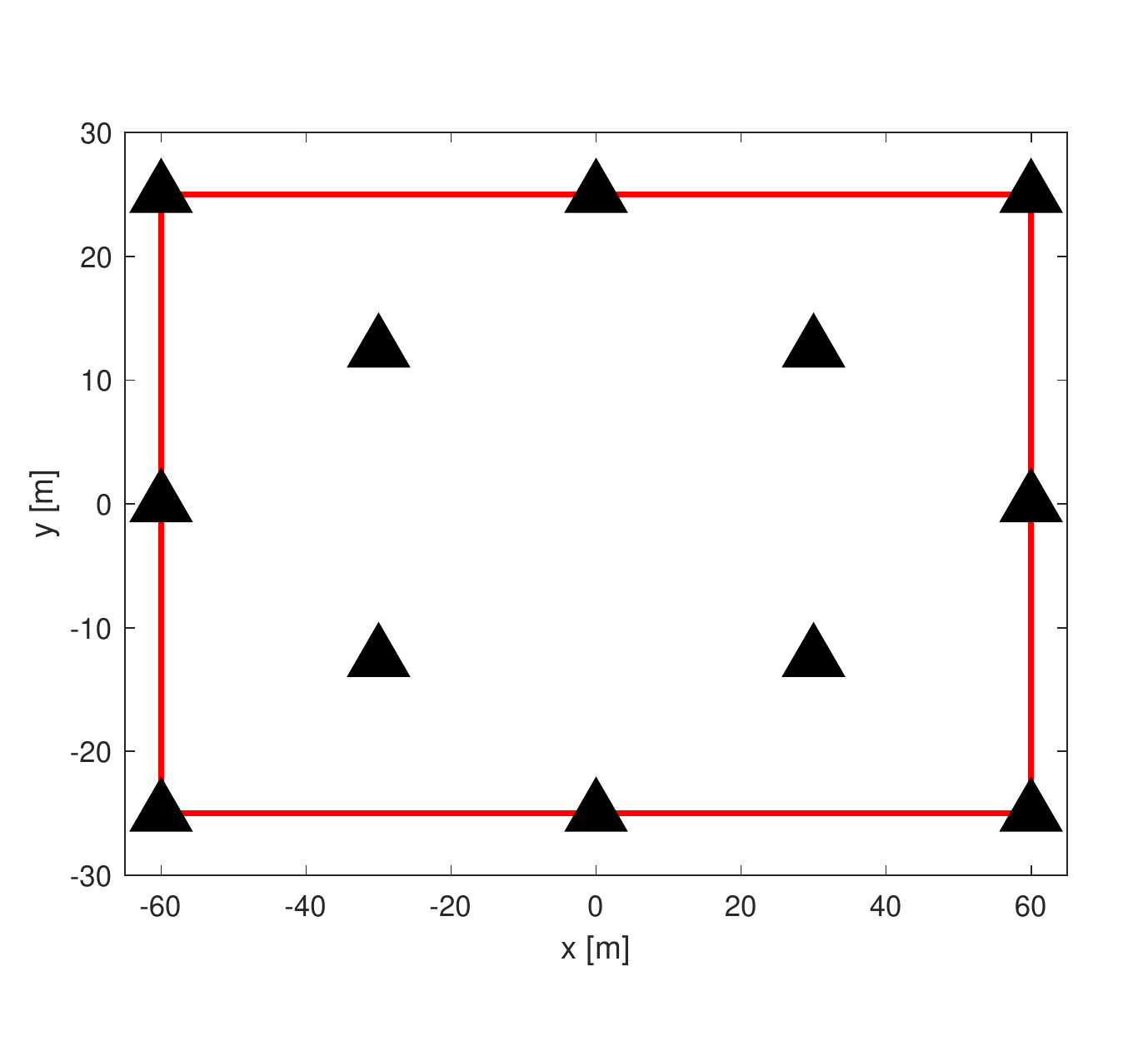}}
	\subfloat[][\label{fig:results:mixed_inf}Mixed deployment for InF.]{\includegraphics[clip=true,trim={3.5cm 8.5cm 3.5cm 8.5cm},width=0.50\linewidth]{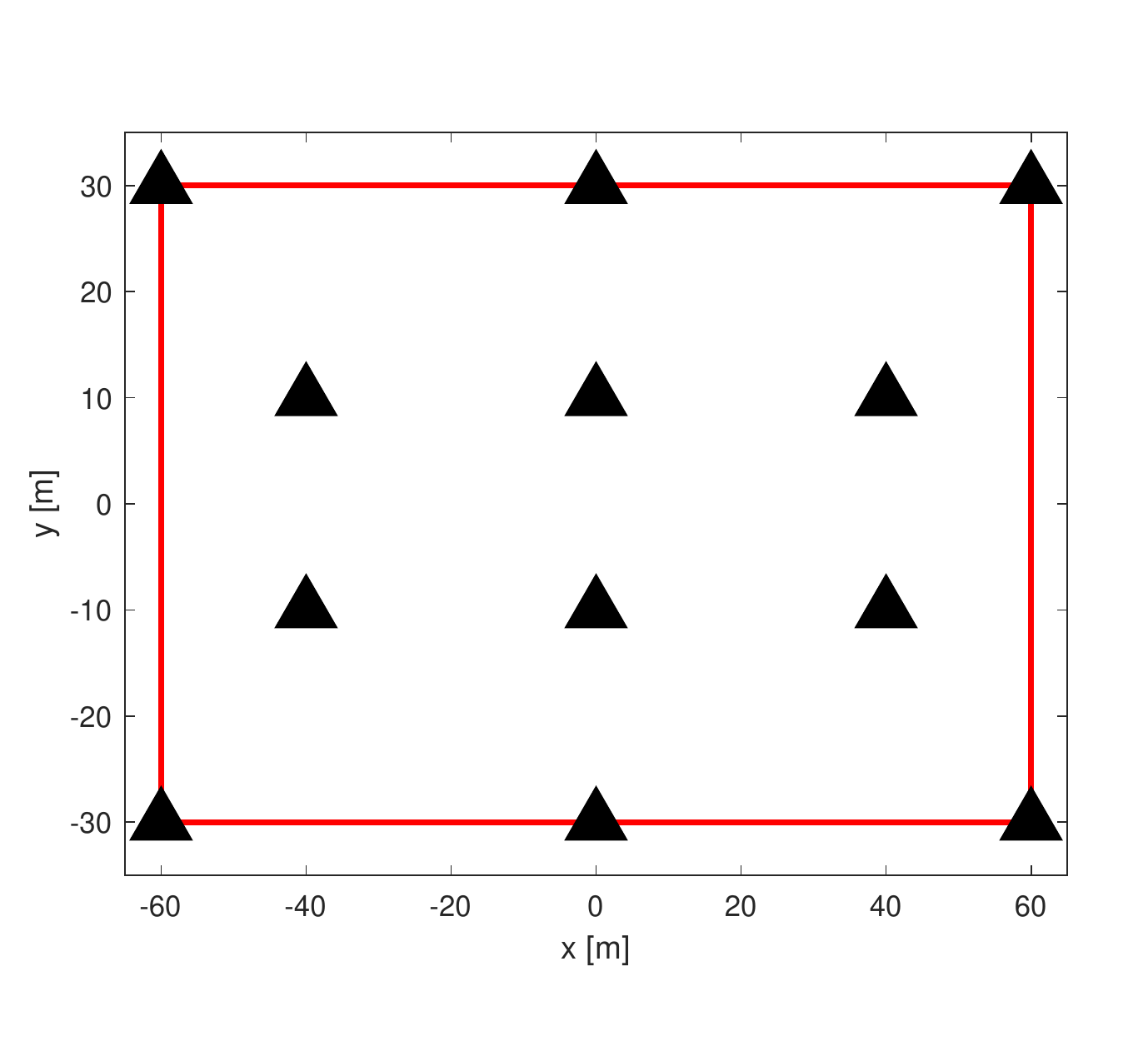}}\\
	\caption{%
		Deployment design strategies.}
	\label{fig:results:deployments}
\end{figure}

\begin{itemize}
	\item \textbf{Standard deployment}: The 3GPP standard deployment. For fair comparison between the deployment strategies in IOO and InF a 12 TRP deployment is considered for both scenarios. In the text to follow, standard deployment refers to standard 12 TRP deployment for both scenarios.
	\item \textbf{Edge deployment}: All TRPs are placed around the edges of the deployment area. In the text to follow, edge deployment refers to 12 TRP (at the edges) deployment for both scenarios.
	\item \textbf{Mixed deployment}:A mix of the standard and edge deployment strategies.
\end{itemize}
The IOO scenario is analyzed with the deployment strategies illustrated in Figures~\ref{fig:results:deployments}(a)--(c) and the InF scenario with the deployment strategies in Figures \ref{fig:results:deployments}(a), (b) and (d). Notice that slightly different mixed deployments are used for IOO and InF (compare Figures \ref{fig:results:deployments}(c) and (d)). The triangles in the figures represent the 5G TRPs and the red lines show the deployment area boundaries. Apart from these, in Section \ref{sec:sim:densification} we analyze the effect of TRP densification in positioning accuracy in IOO scenario.

\section{Performance Evaluation}\label{sec:sim_results}
In this section we show simulation results taking into account realistic channel models that depict typical propagation environments in IOO and InF scenarios. An internal Ericsson simulation tool is used which applies the channel model and parameters agreed by 3GPP in \cite{TR38855,TR38901} to generate DL-TDOA measurements of PRSs between TRPs and UEs as well as LOS information and more. These measurements are then used for positioning estimation. The PRSs are OFDM modulated with parameters are reported in Table \ref{table:common_params} 

\begin{table}
	\caption{Simulation parameters common for IOO and InF.}
	\label{table:common_params}
	\centering
	\begin{tabular}{ |l|c| } 
		\hline
		\textbf{Simulation parameter} & \textbf{Value} \\
		\hline
		\hline
		Carrier frequency & 2~GHz \\
		\hline
		Subcarrier spacing & 30~kHz \\
		\hline
		No of subcarriers & 4096 \\
		\hline
		PRS bandwidth & 100~MHz \\
		\hline
	\end{tabular}
\end{table}

\begin{figure}[ht!]
	\centering
	\subfloat[][\label{fig:results:gdop_s5}GDOP, standard deployment.]{\includegraphics[clip=true,trim={3.5cm 8.5cm 3.5cm 8.5cm},width=0.50\linewidth]{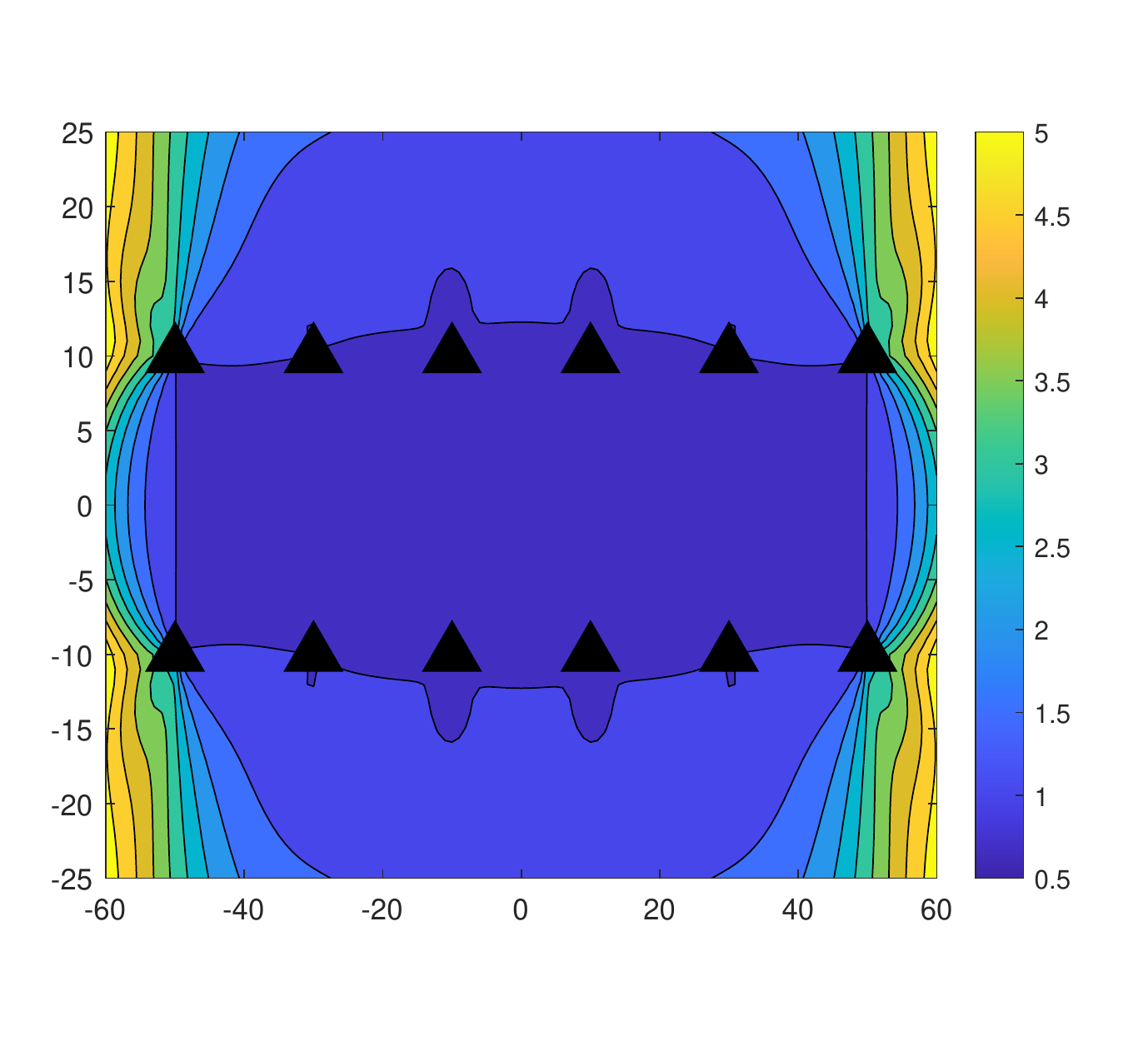}}
	\subfloat[][\label{fig:results:gdop_s6}GDOP edge deployment.]{\includegraphics[clip=true,trim={3.5cm 8.5cm 3.5cm 8.5cm},width=0.50\linewidth]{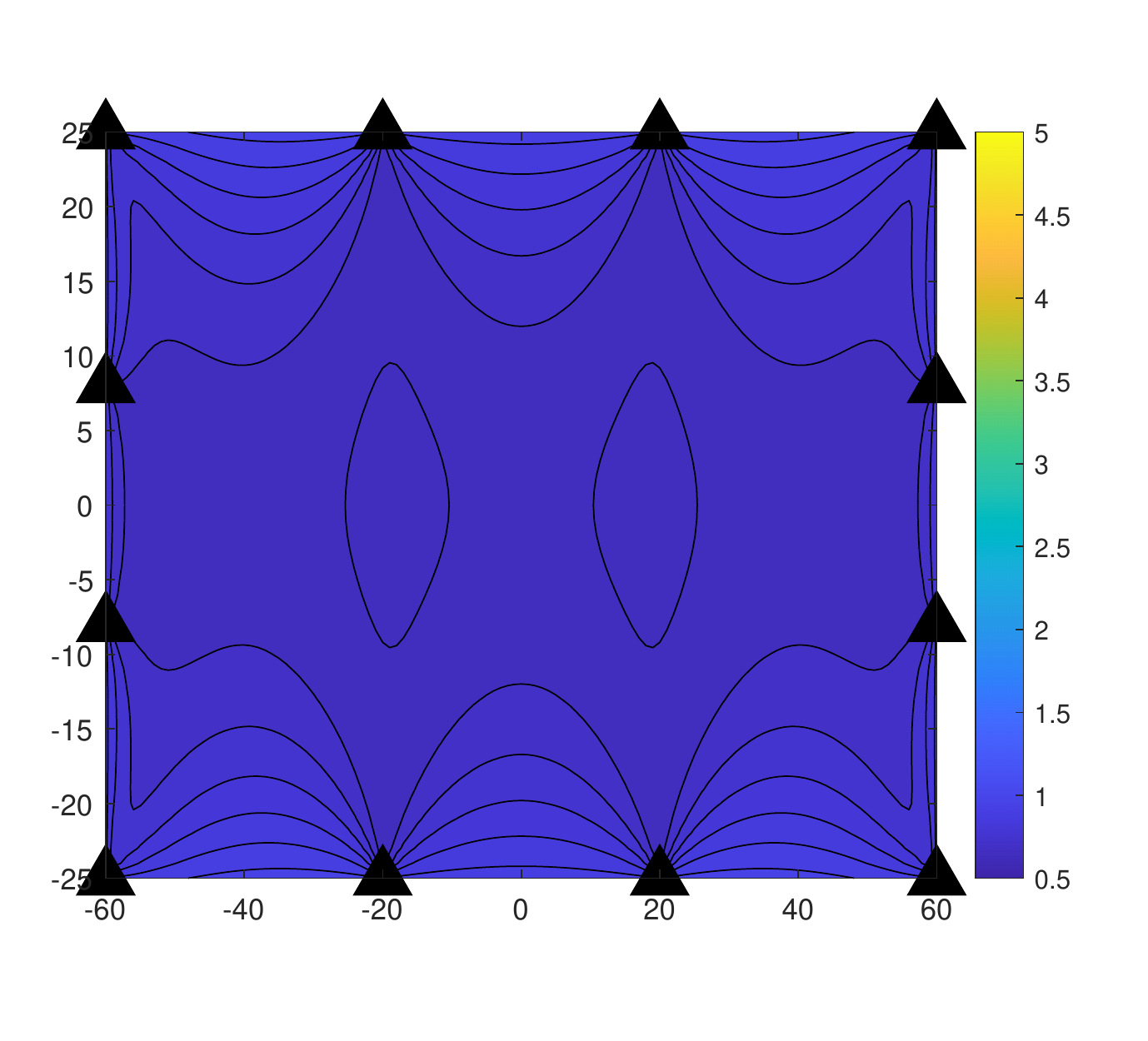}}\\
	\subfloat[][\label{fig:results:crlb_tdoa5}CRLB standard deployment.]{\includegraphics[clip=true,trim={1.5cm 4.5cm 1.5cm 8.5cm},width=0.50\linewidth]{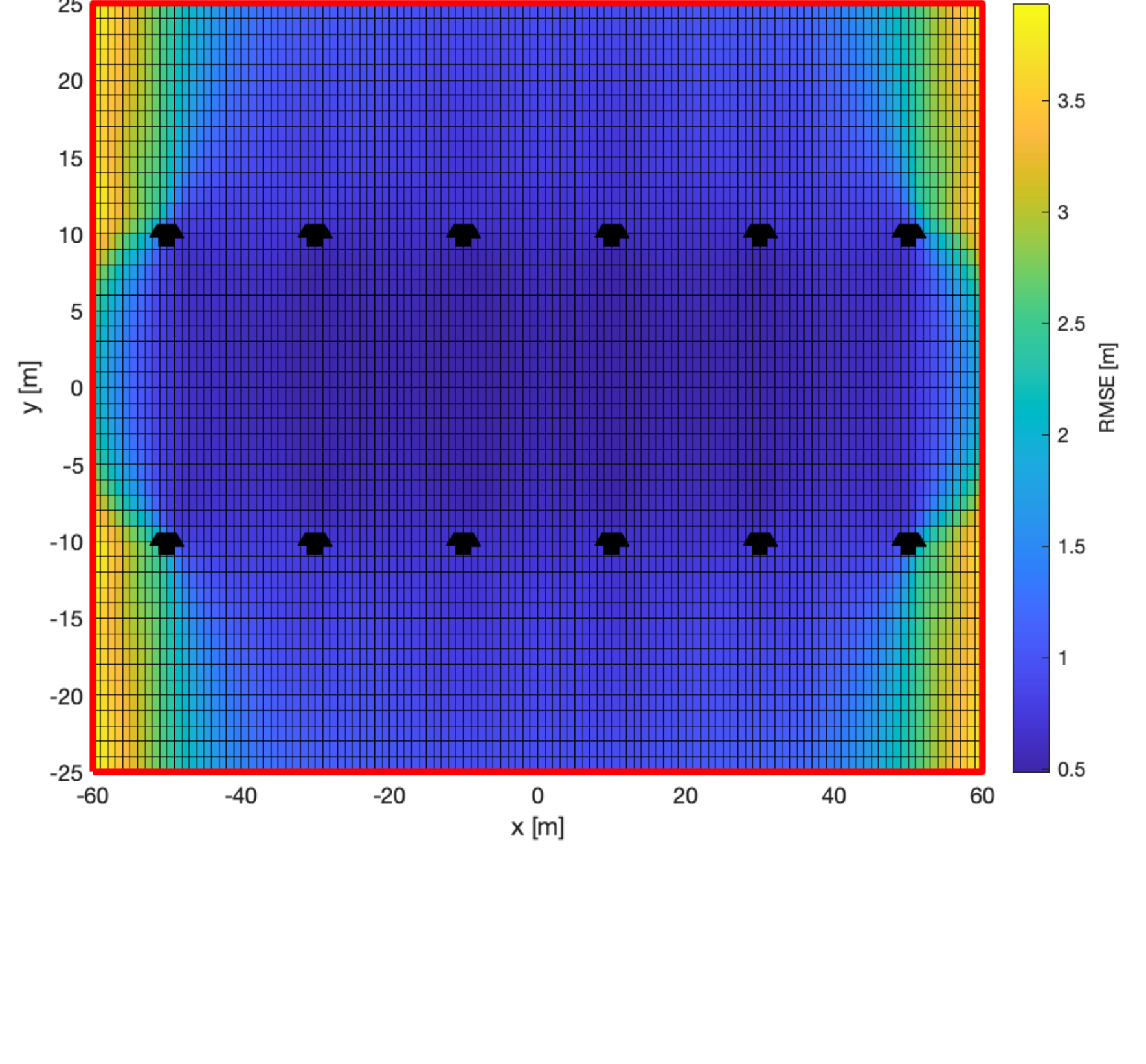}}
    \subfloat[][\label{fig:results:crlb_tdoa6}CRLB edge deployment.]{\includegraphics[clip=true,trim={1.5cm 4.5cm 1.5cm 8.5cm},width=0.50\linewidth]{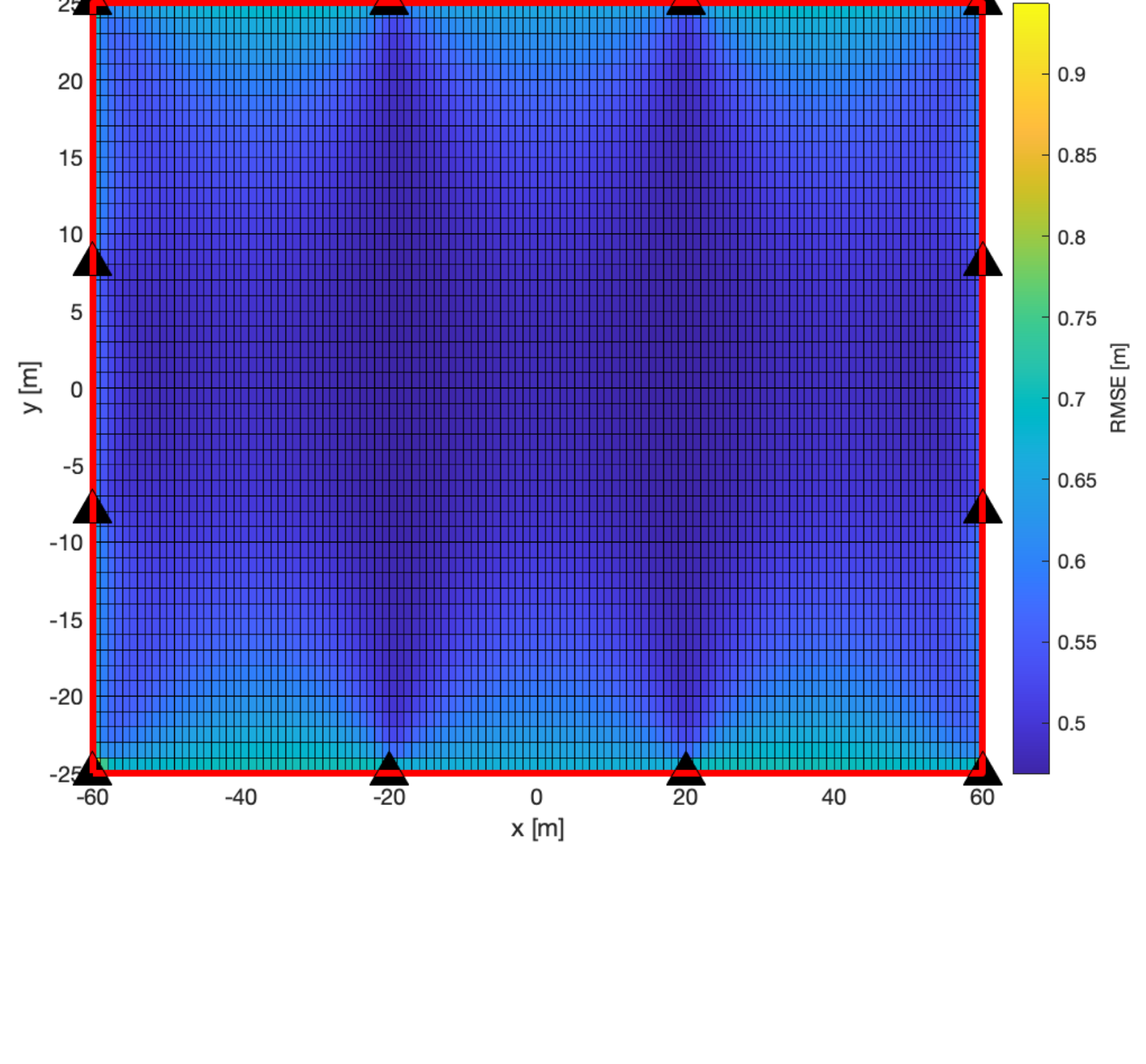}}\\
	\caption{Contour plots showing the GDOP and CRLB in IOO with standard and edge deployments.}
	\label{fig:results:gdop_crlb}
\end{figure}

In Fig.~\ref{fig:results:gdop_crlb}(a) and \ref{fig:results:gdop_crlb}(b), contour plots of the GDOP for the standard and edge deployments in IOO are presented. The GDOP is in general lower and more homogeneous in the edge deployment in comparison to the GDOP when the deployment is based on standard layout. The standard deployment procures worse GDOP at the corners and along the short sides of the deployment area.
The contour plots indicate that the edge deployment is more favorable in terms of receiving high positioning accuracy compared to the standard deployment as the dilution of positioning accuracy due to the geometry of the TRP deployment is lower. 
	
Fig. \ref{fig:results:gdop_crlb}(c) and \ref{fig:results:gdop_crlb}(d) show the CRLB (Equation (\ref{eq:theory:rmse_2d})) of the UE position estimates obtained with DL-TDOA. 
What can be noted is that the CRLB is larger for the standard deployment than for the edge deployment. There are also much larger variations in CRLB for the standard deployment compared to the edge deployment. In the standard deployment, the CRLB range is from 0.49m to 3.94m while in the edge deployment the range is from 0.47m to 0.95m. These observations show that the measurement error is not the only factor that limits the achievable UE localization accuracy. The effect of the geometry of the TRP deployment is also of paramount interest.

\subsection{Indoor Open Office }
NLOS conditions between the TRP $i\in \{1,\dots,N\}$ and the UE adds a positive offset to Equation \eqref{eq:theory:toa} since the signal has traveled longer than the euclidean distance $|\theta - p^i|$. The offset can result in a biased position estimate. 
	
\begin{figure}[ht!]
	\centering
	\includegraphics[width=clip=true,trim={3.5cm 8.5cm 3.5cm 8.5cm},width=1\linewidth]{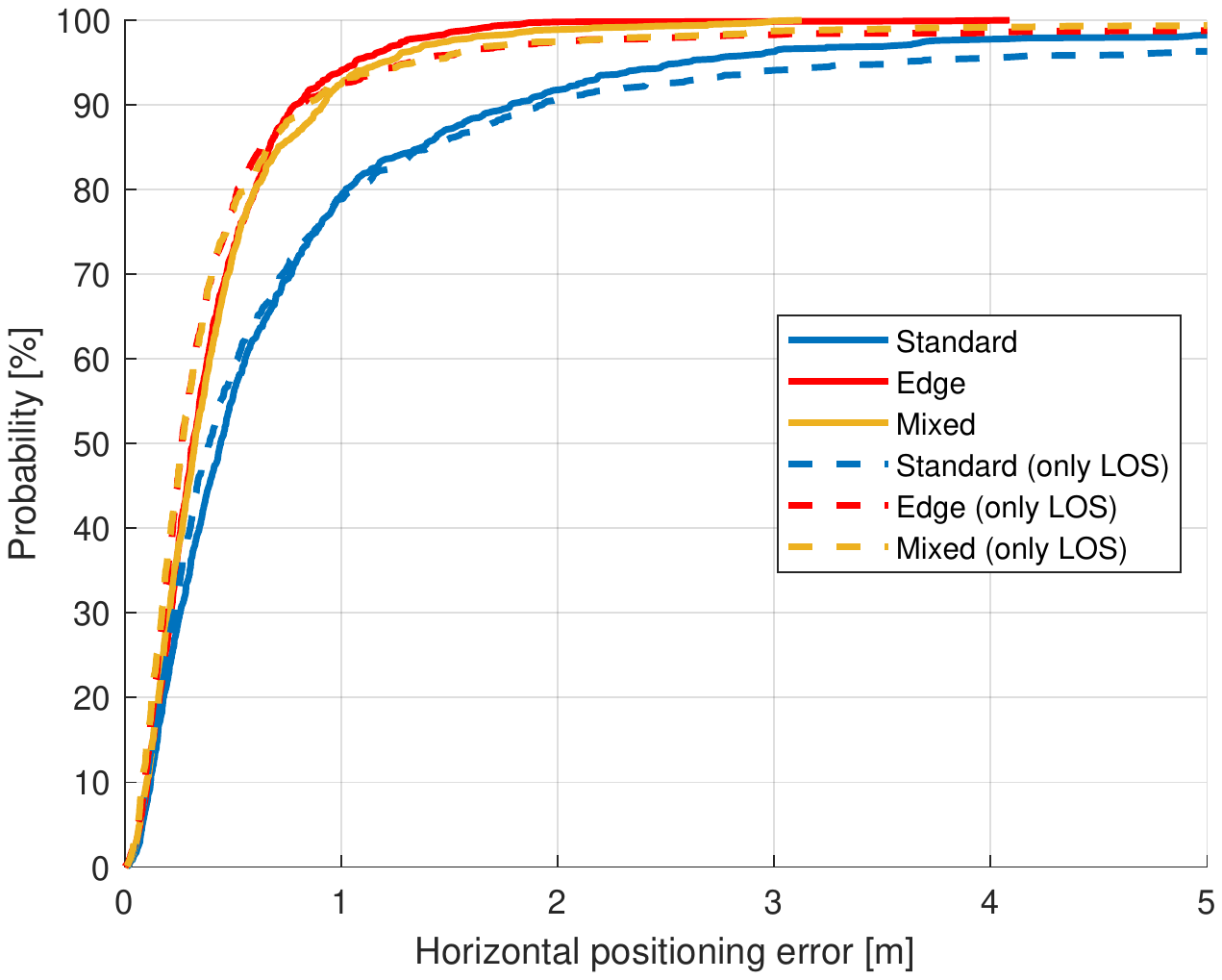}
	\caption{%
		CDFs showing the positioning error in IOO.}
	\label{fig:cdf_ioo_12}
\end{figure}

\begin{figure}[]
	\subfloat[][\label{fig:los_standard_ioo}Standard deployment.]{\includegraphics[clip=true,trim={3.5cm 8.5cm 3.5cm 8.5cm},width=0.50\linewidth]{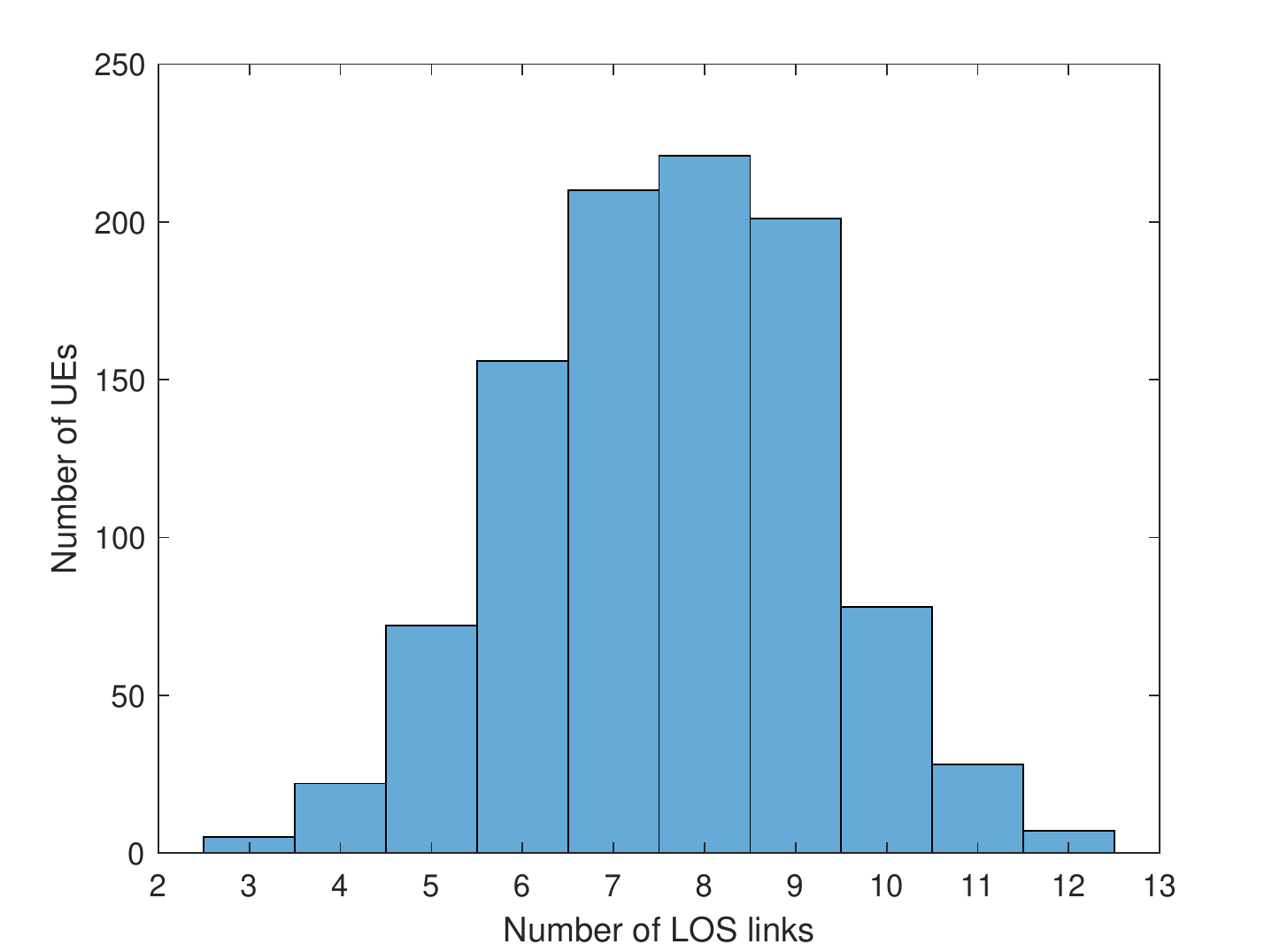}}
	\subfloat[][\label{fig:los_edge_ioo}Edge deployment.]{\includegraphics[clip=true,trim={3.5cm 8.5cm 3.5cm 8.5cm},width=0.50\linewidth]{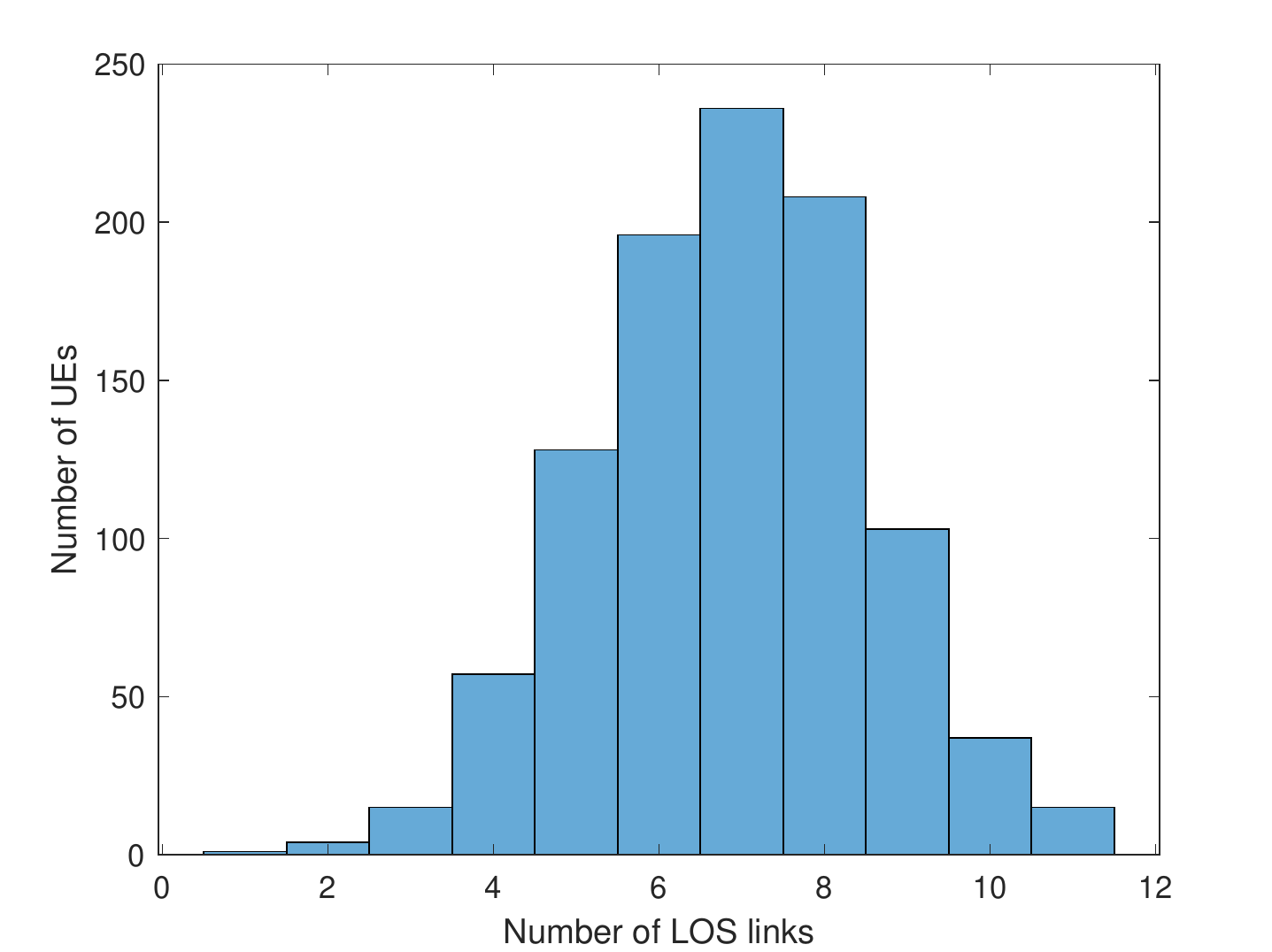}}
	\caption{%
		Histograms showing the number of UEs with a certain number of LOS links to the deployed TRPs for the standard and edge deployments in IOO.}
	\label{fig:los_ioo_12}
\end{figure}

Fig.~\ref{fig:cdf_ioo_12} visualizes the cumulative distribution functions (CDFs) of the positioning error for the standard, edge and mixed deployments in IOO when either all measurements or measurements from only LOS TRPs are used. In this figure we can observe that the edge and the mixed deployments yield similar positioning accuracy and the achievable accuracy is higher than when the TRP geometry is based on standard layout. This result is in line with our theoretical analysis based on GDOP and CRLB. Moreover, the results indicate only small differences in positioning accuracy between either using measurements from all TRPs or using the measurements only corresponding to the TRPs that are in LOS condition with the UE. This can be understood by considering the LOS statistics reported in Fig.~\ref{fig:los_ioo_12}. The histograms show the distribution of the number of LOS TRPs in standard and edge deployment scenarios. It is worth noting that in both scenarios there are very few UEs with less than four LOS links, which is the smallest number of links required to obtain a position estimate using DL-TDOA, and therefore the positioning error distribution looks similar when the UE position is estimated using measurements from all TRPs and using measurements from only LOS TRPs. Moreover, more UEs can maintain a LOS condition when the TRP deployment follows edge layout.

\subsection{Indoor Factory}
\begin{figure}[ht!]
	\centering
	\includegraphics[width=clip=true,trim={3.5cm 8.5cm 3.5cm 8.5cm},width=1\linewidth]{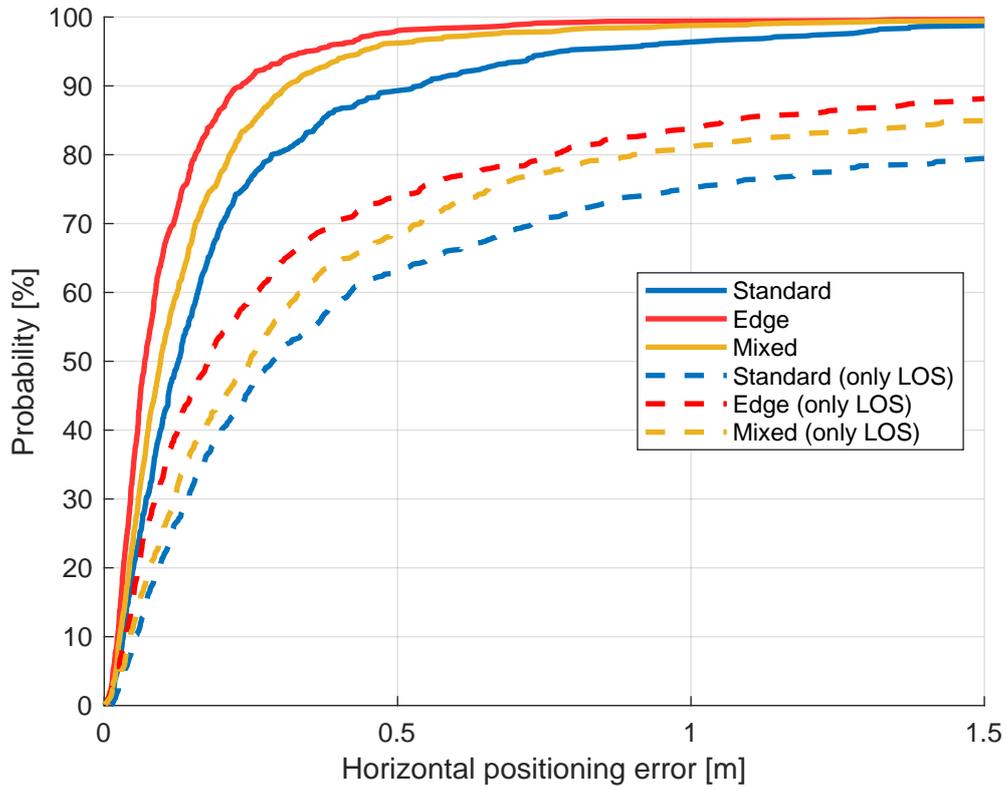}
	\caption{%
		CDFs showing the positioning error in InF-SH.}
	\label{fig:results:cdf_inf_sparse_12}
\end{figure}

\begin{figure}[ht!]
	\subfloat[][\label{fig:results:los_standard_sparse_12}Standard deployment in InF-SH.]{\includegraphics[clip=true,trim={3.5cm 8.5cm 3.5cm 8.5cm},width=0.50\linewidth]{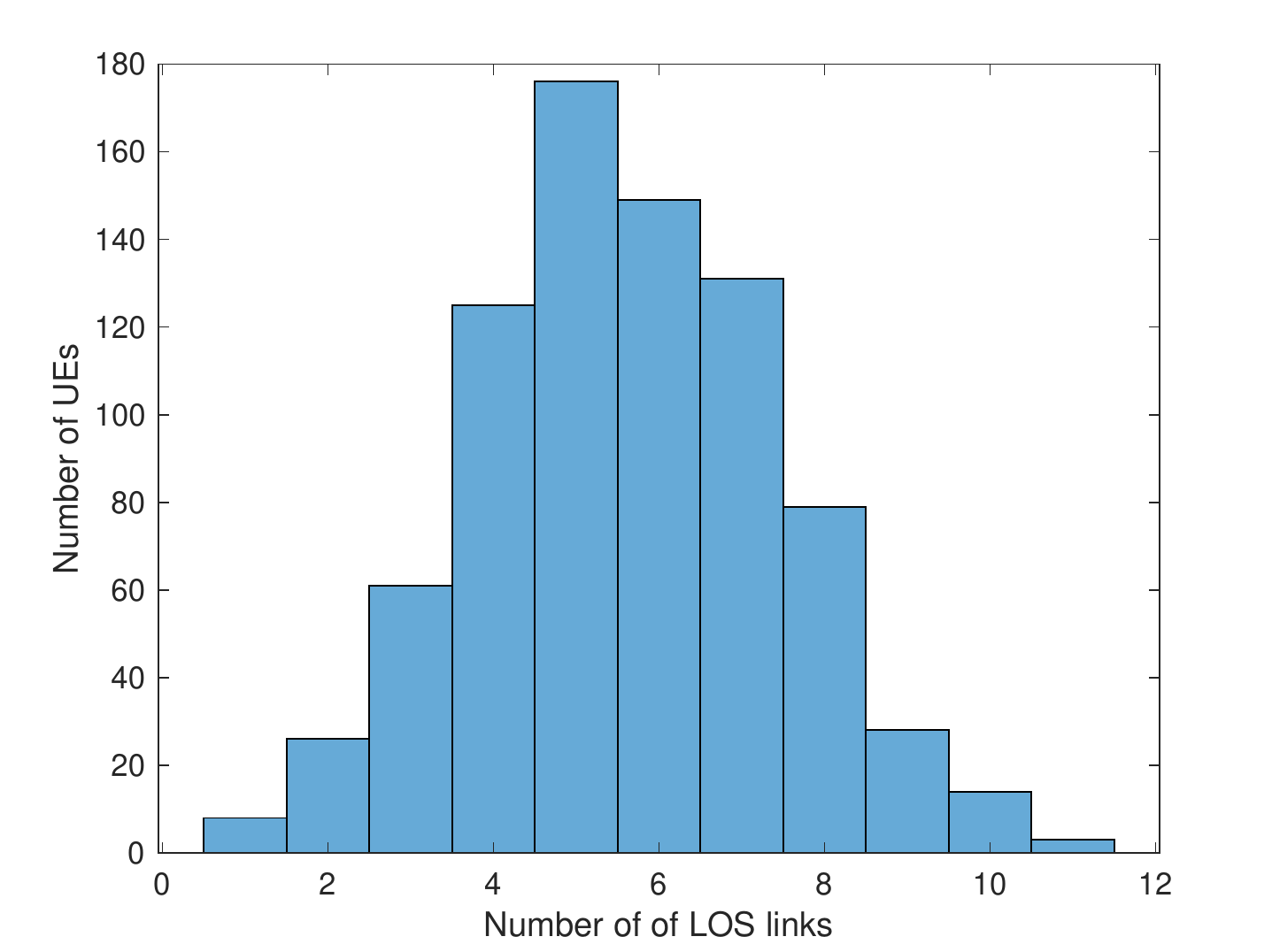}}
	\subfloat[][\label{fig:appendix:los_edge_sparse_12}Edge deployment in InF-SH.]{\includegraphics[clip=true,trim={3.5cm 8.5cm 3.5cm 8.5cm},width=0.50\linewidth]{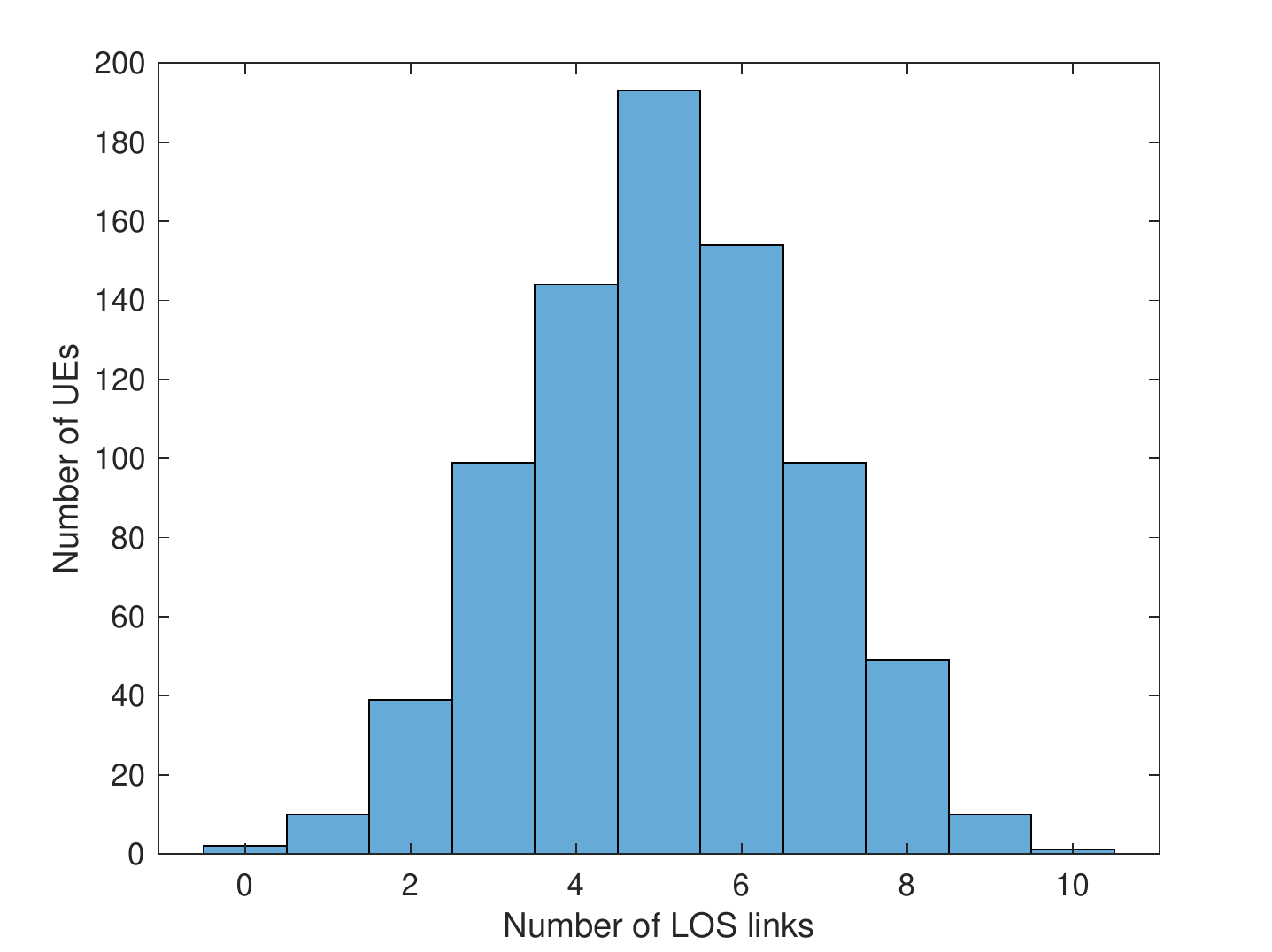}}\\
	\subfloat[][\label{fig:results:los_standard_dense_12}Standard deployment in InF-DH.]{\includegraphics[clip=true,trim={3.5cm 8.5cm 3.5cm 8.5cm},width=0.50\linewidth]{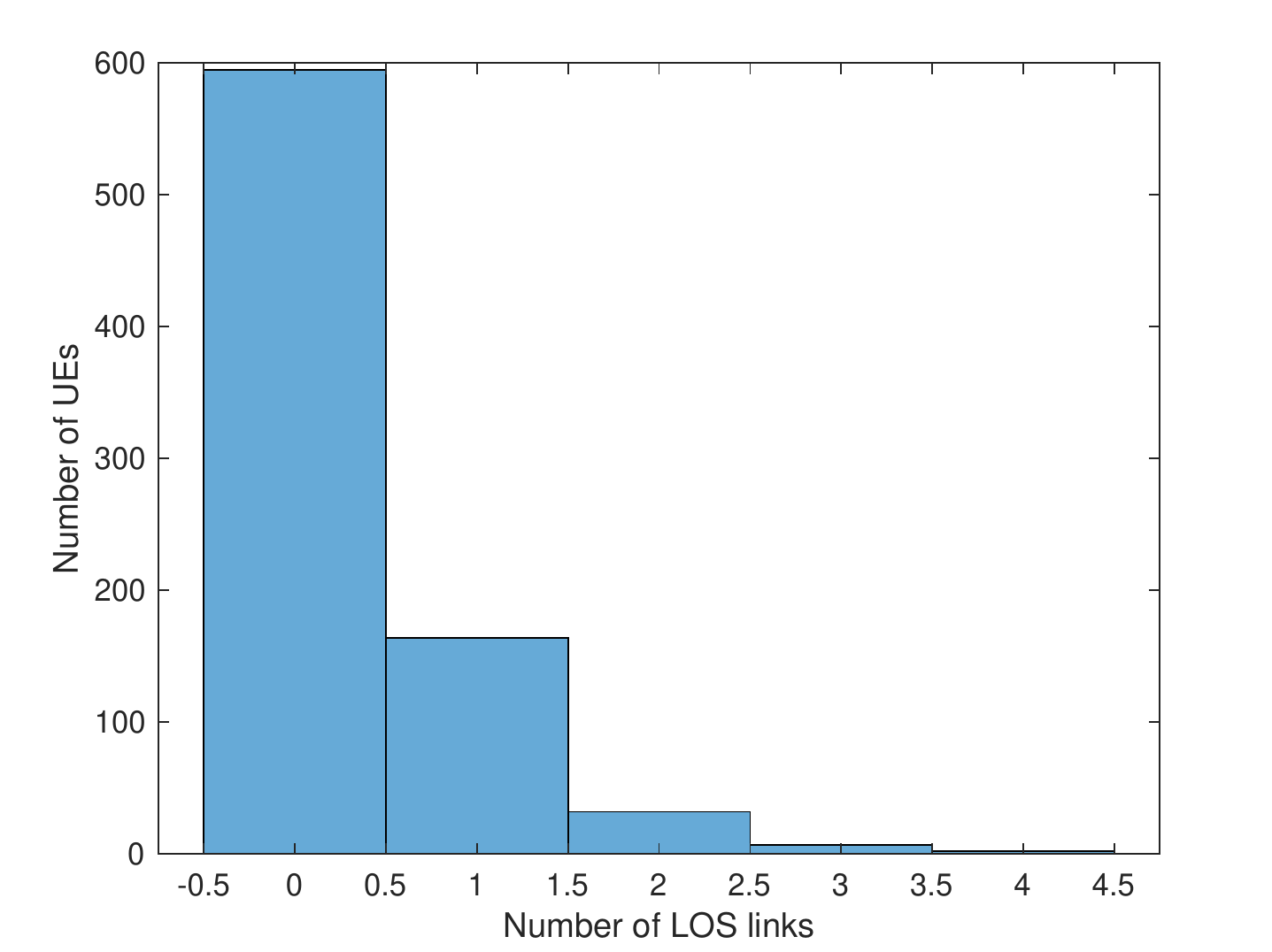}}
	\subfloat[][\label{fig:appendix:los_edge_dense_12}Edge deployment in InF-DH.]{\includegraphics[clip=true,trim={3.5cm 8.5cm 3.5cm 8.5cm},width=0.50\linewidth]{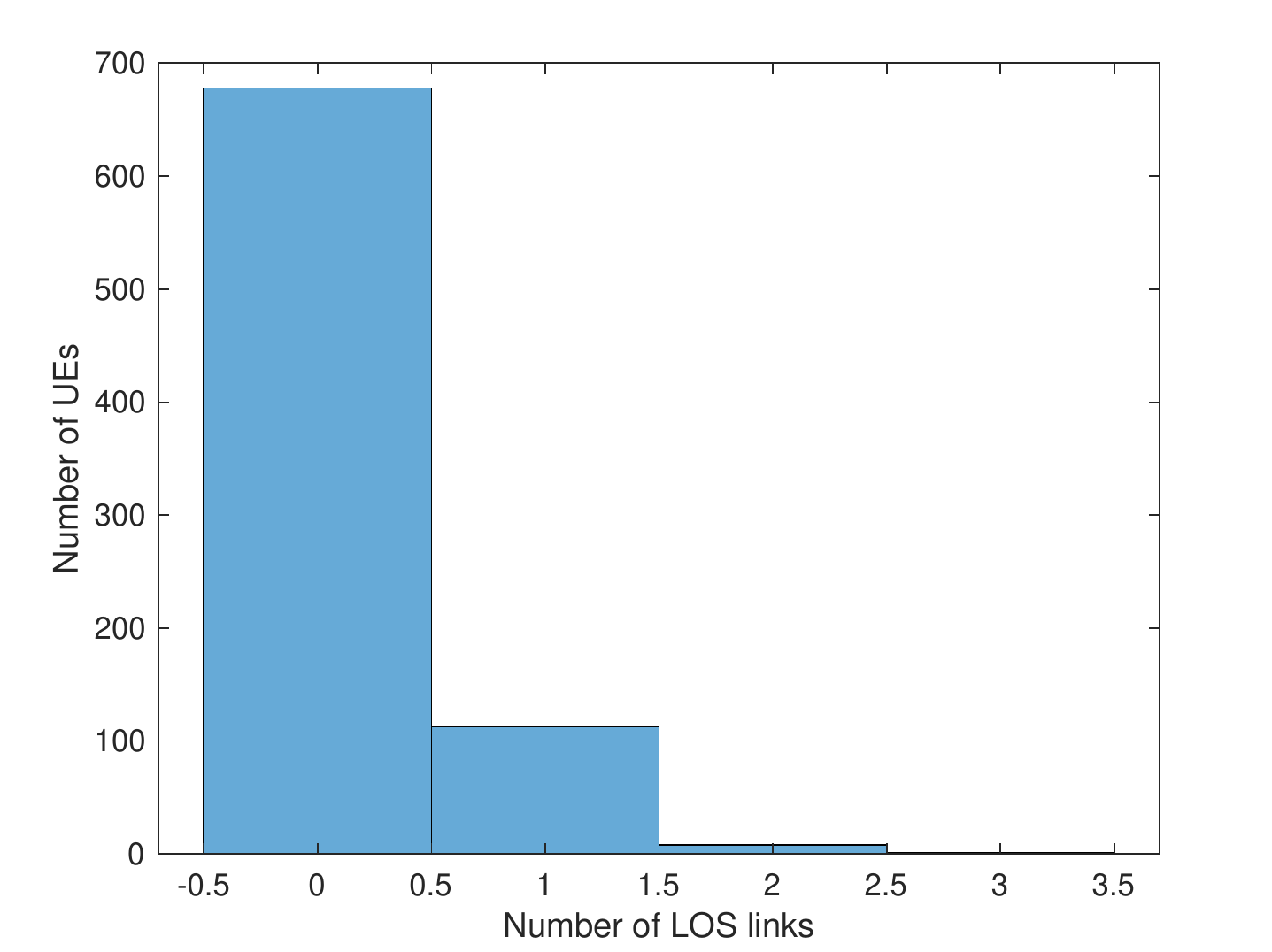}}
	\caption{%
		Histograms showing the number of UEs with a certain number of LOS links to the deployed TRPs for the standard and edge deployments in InF-SH and InF-DH.}
	\label{fig:results:los_inf_12}
\end{figure}

Fig.~\ref{fig:results:cdf_inf_sparse_12} shows the CDFs of the positioning error for the standard, edge and mixed deployments in the InF-SH scenario. It can be observed that in general the achievable positioning accuracy is higher than in IOO scenario and the edge deployment gives better accuracy than the standard layout deployment of the TRPs. The higher accuracy positioning in InF-SH, in comparison to the IOO, is due to higher probability of TRPs being in LOS with the UE. Fig.\ref{fig:results:los_standard_sparse_12} and Fig.\ref{fig:appendix:los_edge_sparse_12} show the LOS statistics for InF-SH standard and InF-SH edge deployments. 
The LOS statistics for InF-DH standard and InF-DH edge deployments are reported in Fig. \ref{fig:results:los_standard_dense_12} and Fig.\ref{fig:appendix:los_edge_dense_12}. With majority of UEs having no LOS link with any of the deployed TRPs InF-DH, where more than 40\% of the area is covered by clutters, is a challenging environment for precise UE positioning. In the sub-section to follow we address the problem of enhancing positioning accuracy in InF-DH scenario by densification of TRP. Moreover, the effect of TRP densification on achievable positioning accuracy in IOO scenario is also evaluated.

\subsection{TRP Densification}
\label{sec:sim:densification}
\begin{figure}[ht!]
	\centering
	\subfloat[][\label{fig:percentiles_standard}Standard.]{\includegraphics[clip=true,trim={3.5cm 8.5cm 3.5cm 8.5cm},width=0.50\linewidth]{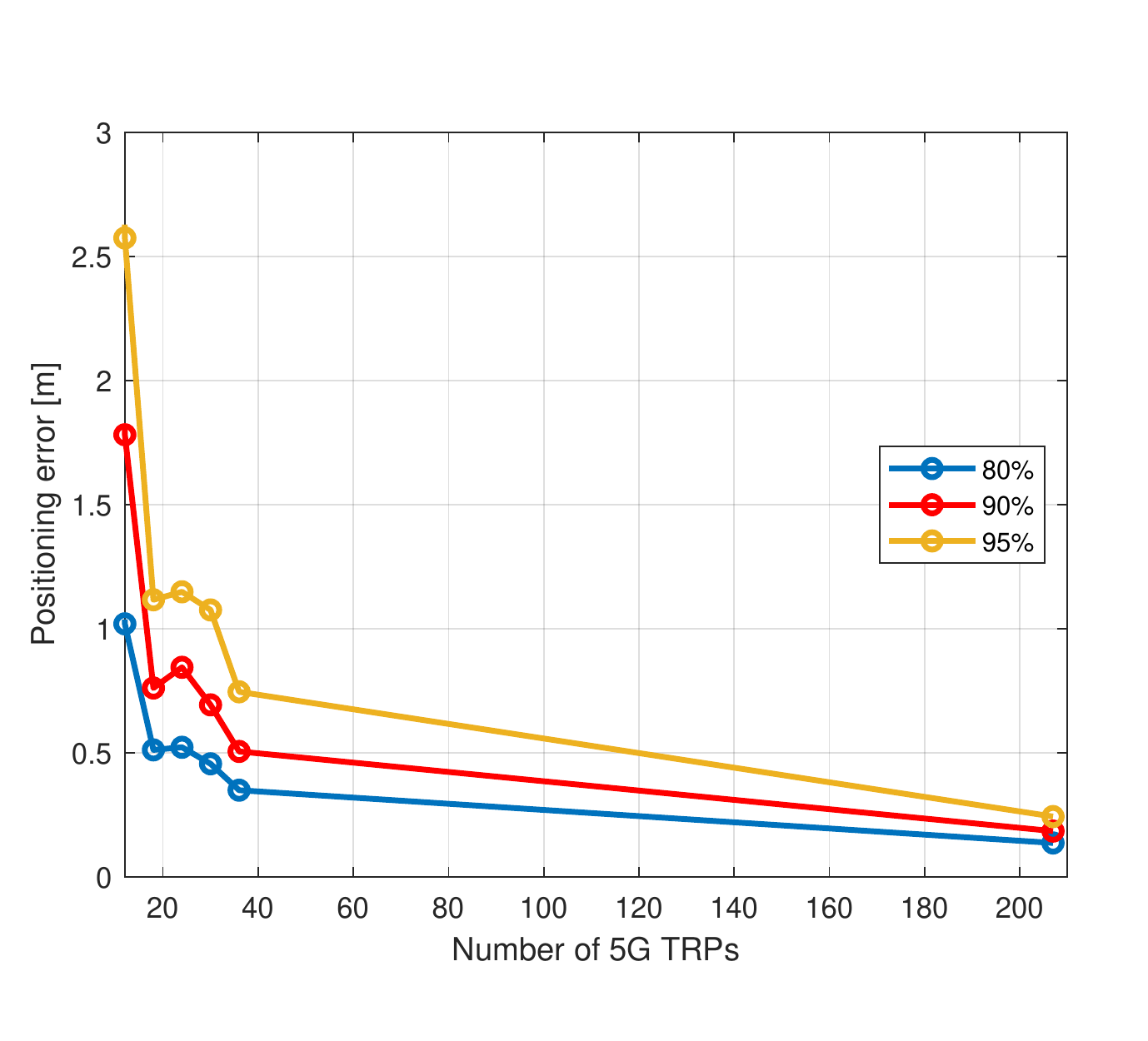}}
	\subfloat[][\label{fig:percentiles_edge}Edge.]{\includegraphics[clip=true,trim={3.5cm 8.5cm 3.5cm 8.5cm},width=0.50\linewidth]{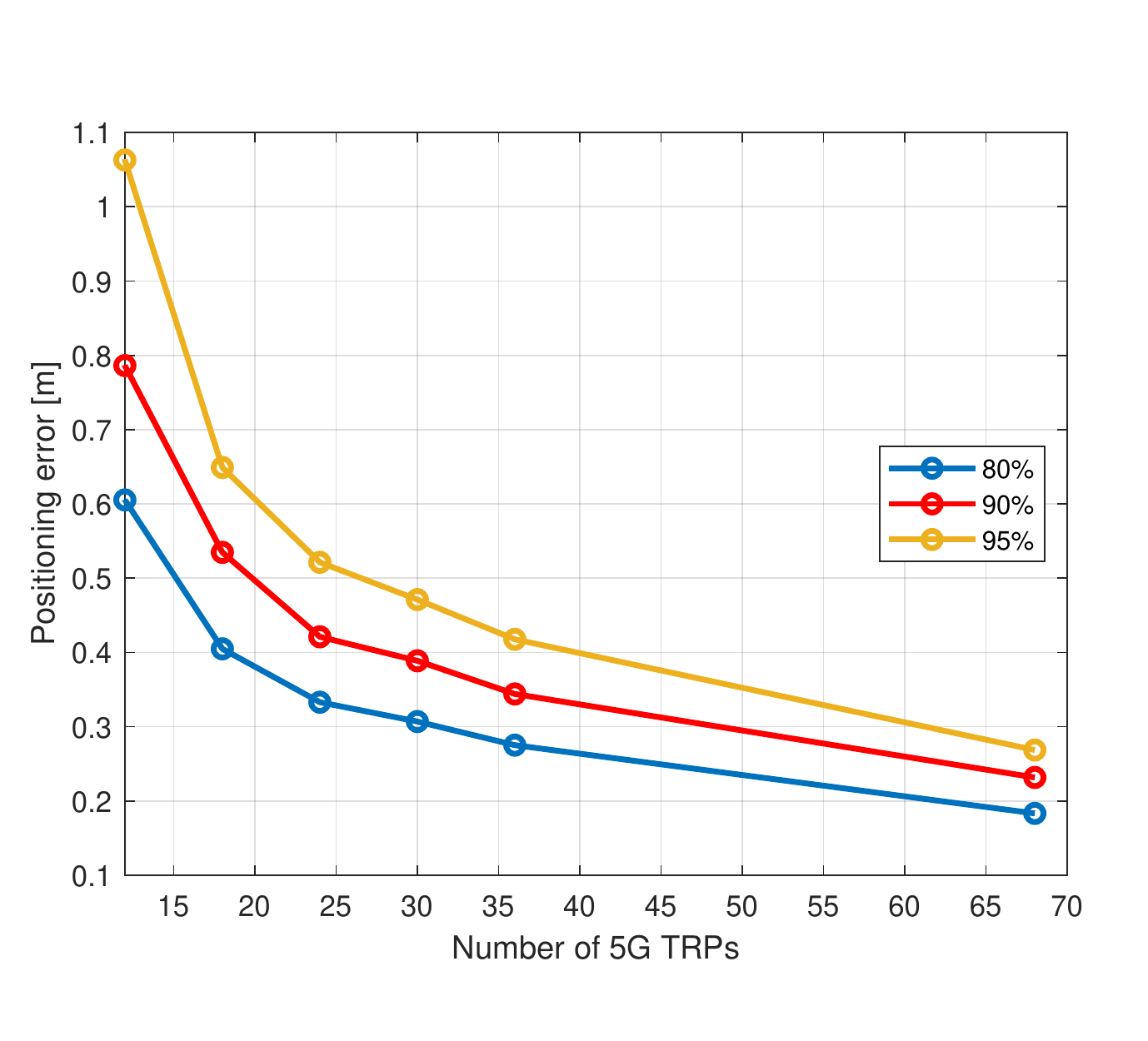}}\\
	\caption{%
		Curves showing the change in positioning accuracy at different percentiles when the number of TRPs increases in standard and edge deployments in IOO.}
	\label{fig:percentiles}
\end{figure}

The effect of TRP densification on achievable positioning accuracy in IOO scenario (both standard and edge) is shown in Fig.\ref{fig:percentiles}. It has been observed that the positioning accuracy improves when the number of TRP is increased. When the number of TRP, following the standard deployment layout, is increased beyond 36 the achievable positioning accuracy tends to saturate. Similar observation can also be made when the number of TRP is increased following the edge deployment layout. These observations validate that in either of the deployment strategies the improvement in positioning accuracy is limited by the number of TRP that can be deployed in a scenario like IOO.

\begin{figure}[ht!]
	\centering
	\subfloat[][\label{fig:results:worst_ues_dense_mixed_12}Original mixed deployment with a positioning error of 13.11~m at the 90 percentile.]{\includegraphics[clip=true,trim={3.5cm 8.5cm 3.5cm 8.5cm},width=0.49\linewidth]{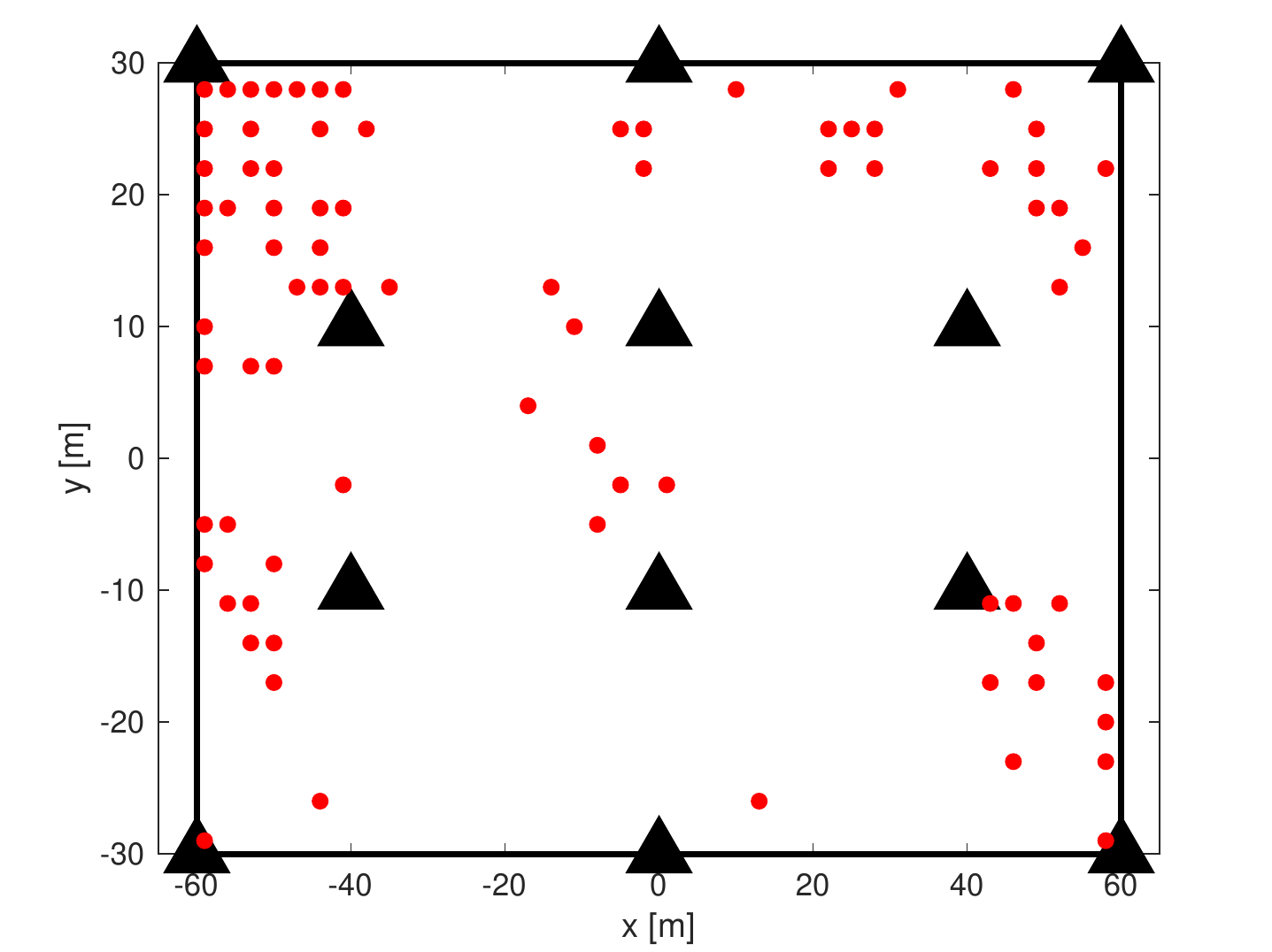}}
	\hfill
	\subfloat[][\label{fig:results:worst_ues_dense_mixed_13}Mixed deployment with one additional TRP and a positioning error of 12.13~m at the 90 percentile.]{\includegraphics[clip=true,trim={3.5cm 8.5cm 3.5cm 8.5cm},width=0.49\linewidth]{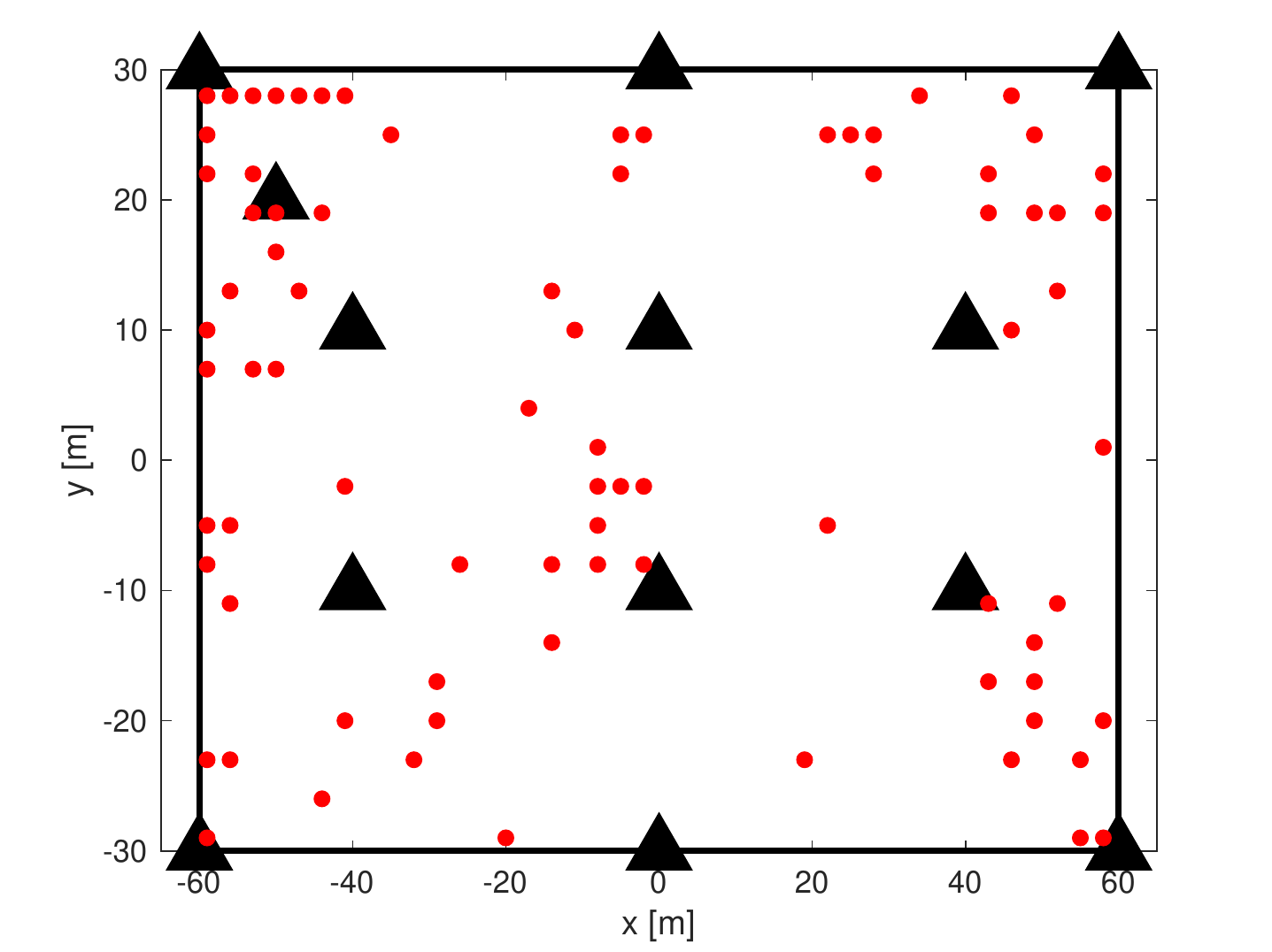}}\\
	\subfloat[][\label{fig:results:worst_ues_dense_mixed_14}Mixed deployment with two additional TRPs and a positioning error of 11.80~m at the 90 percentile.]{\includegraphics[clip=true,trim={3.5cm 8.5cm 3.5cm 8.5cm},width=0.49\linewidth]{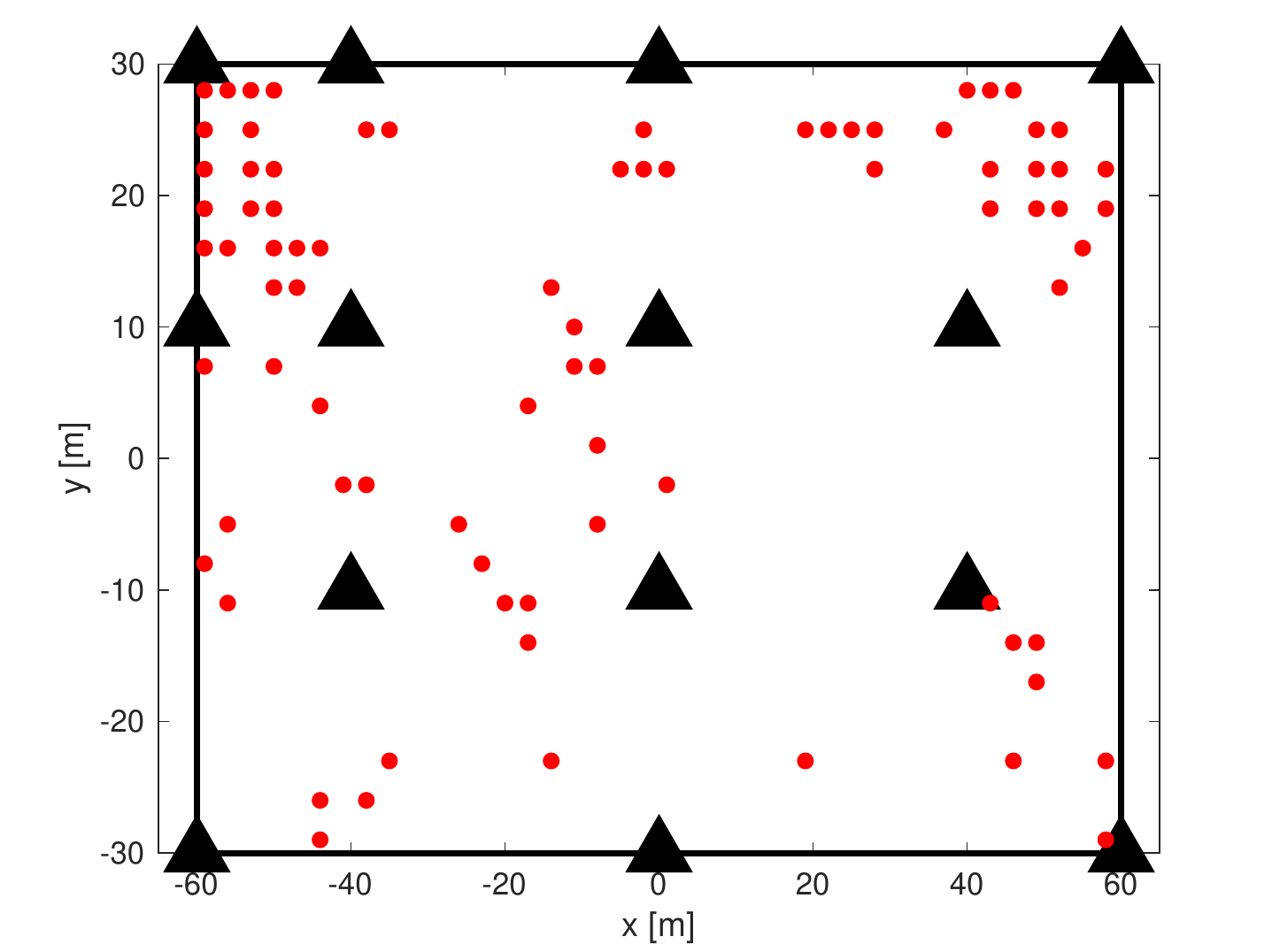}}
	\hfill
	\subfloat[][\label{fig:results:worst_ues_dense_mixed_15}Mixed deployment with three additional TRPs and a positioning error of 11.55~m at the 90 percentile.]{\includegraphics[clip=true,trim={3.5cm 8.5cm 3.5cm 8.5cm},width=0.49\linewidth]{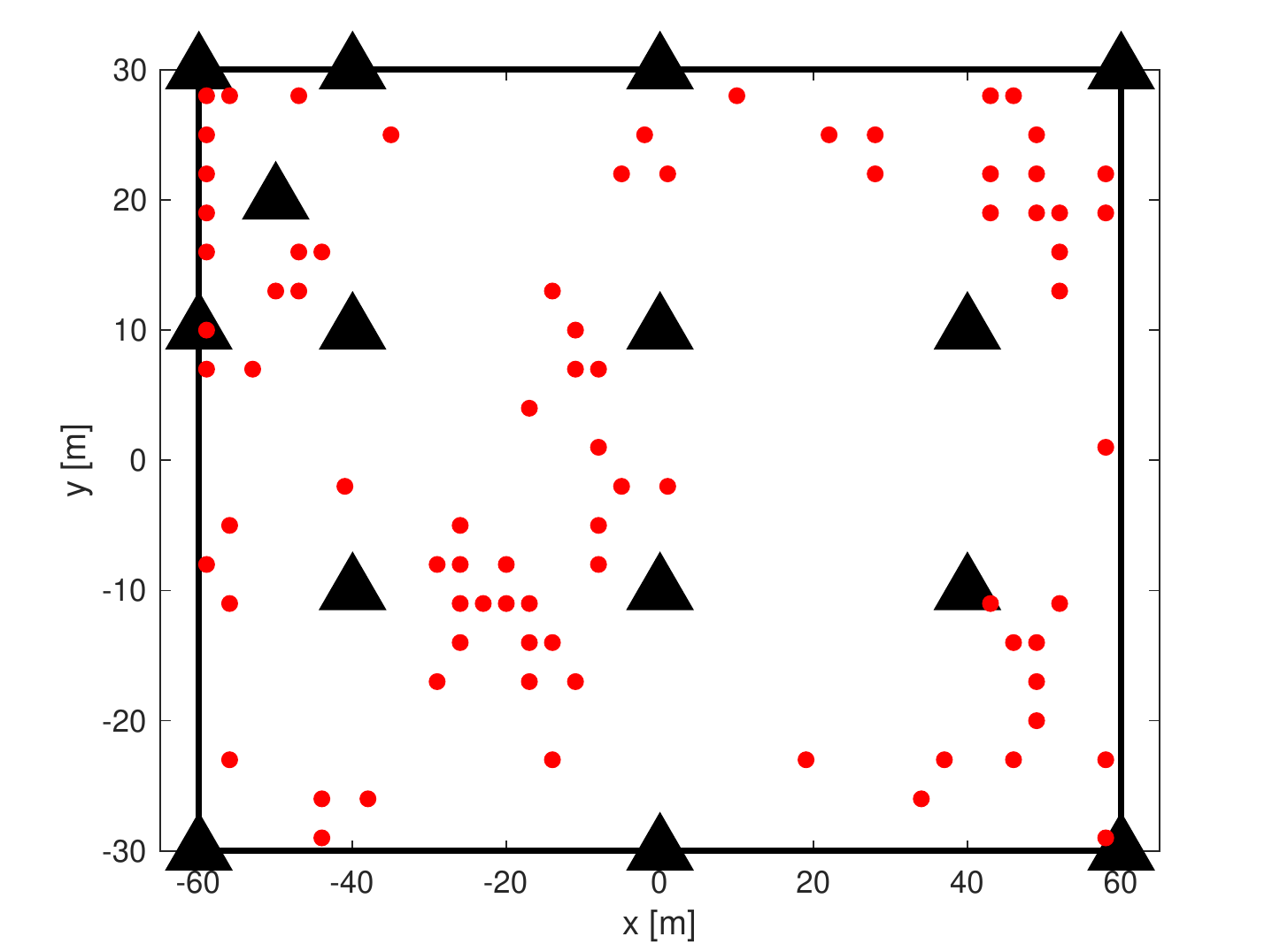}}\\
	\caption{%
		Plots showing the positions of the worst 10\% of the UEs with respect to positioning error for the mixed deployment and the modifications where 1--3 TRPs are added in InF-DH.}
	\label{fig:results:worst_ues_mixed_more_bs}
\end{figure}

\begin{figure}[ht!]
	\centering
	\includegraphics[width=clip=true,trim={3.5cm 8.5cm 3.5cm 8.5cm},width=0.9\linewidth]{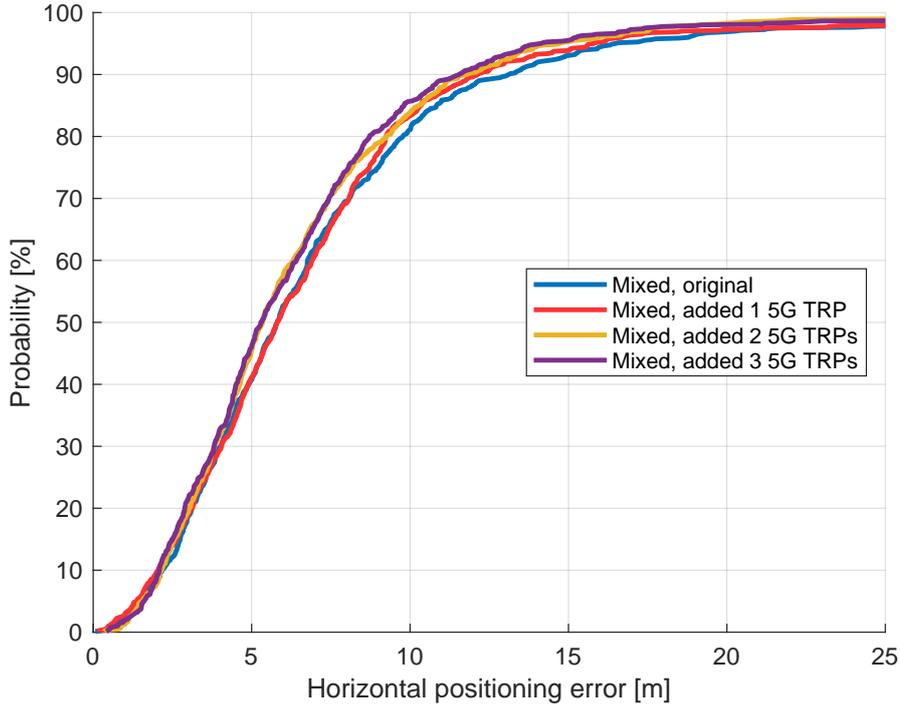}
	\caption{%
		Positioning error CDFs for the mixed InF-DH  deployment when 1-3 TRPs are added.}
	\label{fig:results:cdf_moving_bs}
\end{figure}

Densification of a given deployment can alternatively be done by successive addition of TRPs in areas where the accuracy is bad.
Fig.~\ref{fig:results:worst_ues_mixed_more_bs}(a) shows the location of UEs with the largest positioning errors in InF-DH scenario when the deployment is based on mixed layout. It has been observed that the successive addition of TRPs help improving the achievable positioning accuracy. In particular the 90\% positioning accuracy improves from 13.11m to 12.13m when one additional TRP is deployed in the area where the positioning accuracy is bad. Deploying two more TRPs to the problematic area helps improving the positioning accuracy from 13.11m to 11.80m. Furthermore, the achievable positioning accuracy is 11.55m when three additional TRPs are deployed in the problematic area as shown in Fig.~\ref{fig:results:worst_ues_mixed_more_bs}(b)--(d).
The CDF curves showing the positioning error in the InF-DH scenario when the deployment strategy follows a mixed layout is shown in Fig.~ \ref{fig:results:cdf_moving_bs} where it can be observed that the overall positioning accuracy only improves slightly when TRPs are added.
\subsection{Evaluation summary} \label{sec:perf_eval}

\begin{table}[]
	\caption{Positioning percentile errors for IOO deployment strategies.}
	\centering
	\label{table:results_ioo}
	\begin{tabular}{ *{4}{c|} }	
		\cline{2-4}
		\multirow{2}{4em}{} & \multicolumn{3}{c|}{\textbf{Positioning error [m]}} \\
		\hline
		\multicolumn{1}{|c|}{\textbf{IOO deployment}} &  \underline{80\%}  &   \underline{90\%}  &   \underline{95\%} \\
		\cline{1-1}
		\multicolumn{1}{|c|}{Standard} & 1.02 & 1.78 & 2.58 \\
		\multicolumn{1}{|c|}{Edge} & 0.61 & 0.79 & 1.06 \\
		\multicolumn{1}{|c|}{Mixed} & 0.60 & 0.92 & 1.19 \\
		\hline
	\end{tabular}
\end{table}

\begin{table}[]
	\caption{Positioning percentile errors for InF deployment strategies.}
	\centering
	\label{table:results_inf}
	\begin{tabular}{ *{7}{c|} }
	\cline{2-7}
		\multirow{3}{4em}{} & \multicolumn{6}{c|}{\textbf{Positioning error [m]}} \\
		\cline{2-7}
		& \multicolumn{3}{c|}{\textbf{InF-SH}}
		& \multicolumn{3}{c|}{\textbf{InF-DH}} \\
		\hline
		\multicolumn{1}{|c|}{\textbf{InF deployment}} &   \underline{80\%}  &   \underline{90\%}  &   \underline{95\%}  &   \underline{80\%}  & \underline{90\%}  &   \underline{95\%}\\
		\cline{1-1}
		\multicolumn{1}{|c|}{Standard} & 0.29 & 0.54 & 0.78 & 15.59 & 25.83 & 46.47 \\
		\multicolumn{1}{|c|}{Edge} & 0.16 & 0.23 & 0.34 & 10.18 & 13.56 & 18.04 \\
		\multicolumn{1}{|c|}{Mixed} & 0.21 & 0.31 & 0.44 & 9.77 & 13.11 & 16.48 \\
		\hline
	\end{tabular}
\end{table}

In this paper we presented our study on achievable positioning accuracy in IOO and variety of InF scenarios where the focus is to verify the effect of deployment strategies and TRP densification on DL-TDOA positioning technique. In IOO scenario, we observed effects of various strength coming from the number of TRPs in the deployment, geometry of the deployment, and LOS conditions that affect the achievable positioning accuracy. The observations can be summarized as:
\begin{itemize}
    \item An increasing number of TRPs proves to always improve the positioning accuracy and also the availability.
    \item The deployment geometry is of high importance and must be considered as one of the important factors for high accuracy UE localization.
    \item Better LOS conditions are favorable from a positioning perspective and improve the positioning accuracy. 
\end{itemize}

The study of the two InF scenarios highlighted the difficulty in achieving high accuracy and availability indoor positioning in a densely cluttered deployment area in comparison to a sparsely cluttered deployment area. The vital factor was an immense lack of the number of LOS links which proved that decent LOS conditions are necessary for accurate positioning. The most prominent difference between IOO and InF is that the clutter in InF, especially InF-DH, leads to a great loss in LOS links. This makes the trade-off between favorable geometry and optimizing the LOS links more difficult in InF-DH than in IOO, where the geometry has the biggest impact on the results. Table \ref{table:results_ioo} and Table \ref{table:results_inf} summarizes the achievable positioning accuracy given that the deployment is adjusted based on studied strategies in the studied scenarios.

\section{Conclusions} \label{sec:conclusions}
In this paper, we investigated different aspects of deployment and its impact on 5G indoor positioning. While the standard deployment designs proposed by 3GPP always considers the TRPs to be in the ceiling distributed evenly inside the indoor hall, our studies show that for improved positioning performance, it is more favorable to mount the TRPs evenly on the wall at the edges of the deployment area. In a more reasonable approach, in which both positioning and communication performance requirements are considered, the strategy would be to have a mixed deployment in which most of the TRPs are on the walls and few distributed on the ceiling if the area of deployment is specifically cluttered due to the presence of machines and other objects in a factory floor. A final interesting finding is that if one has to choose between two deployments, one promoting deployment geometry and one promoting the number of LOS links, the choice should fall on the first option.

\begin{acknowledgments}
This work has been supported in parts by Ericsson Research AB and by the European Union’s Horizon 2020 research and innovation programme under grant agreement No.871249 (Research and Innovation Action), LOCalization and analytics on-demand embedded in the 5G ecosystem for Ubiquitous vertical applicationS (LOCUS).
\end{acknowledgments}


\begin{thebibliography}{00}
	
\bibitem{RP-181399}
3GPP RP-181399, Study on NR positioning support, June 2018.

\bibitem{TR38855}
3GPP TR 38.855, Study on NR positioning support, Rel.16. 

\bibitem{TR38901} 3GPP TR 38.901. Study on channel model for frequencies from 0.5 to 100 GHz. Technical report, December 2019.

\bibitem{FCC}
FCC 15-9, \emph{Wireless E911 Location Accuracy Requirements}, fourth report and order, January 2015.

\bibitem{5GSmart} J. Sachs, K. Wallstedt, F. Alriksson and G. Eneroth. 5G and Smart Manufacturing, Ericsson Technology Review, Feb. 2019.

\bibitem{PastPresentFuture} S. M. Razavi, F. Gunnarsson, H. Rydén, Å. Busin, X. Lin, X. Zhang, S. Dwivedi, I. Siomina, and R. Shreevastav. Positioning in Cellular Networks: Past, Present, Future. In 2018 IEEE Wireless Communications and Networking Conference (WCNC), pages 1–6, Apr. 2018. 

\bibitem{Henrik}
H.~Ryd\'{e}n, S. M.~Razavi, F.~Gunnarsson, S. M.~Kim, M.~Wang,
Y.~Blankenship, A.~Gr\"{o}vlen, and A.~Busin. \emph{Baseline Performance of LTE Positioning in 3GPP
	3D MIMO Indoor User Scenarios}, in Proc. International Conf. on Localization and GNSS (ICL-GNSS), 2015.

\bibitem{delPeral2017} J. A. del Peral-Rosado, R. Raulefs, J. A. López-Salcedo, and G. Seco-Granados. Survey of Cellular Mobile Radio Localization Methods: From 1G to 5G. IEEE Communications Surveys Tutorials, 20(2):1124–1148, Secondquarter 2018. 

\bibitem{AccPosUWB} H. Soganci, S. Gezici, and H. V. Poor. Accurate Positioning in Ultra-Wideband Systems. IEEE Wireless Communications, 18(2):19–27, April 2011. 

\bibitem{thesis} J. Ahlander and M. Posluk, ``Deployment Strategies for High Accuracy and Availability Indoor Positioning with 5G'', Master’s thesis, Dept. Elect. Eng., Linköping University, Linköping, Sweden, Jun. 2020. Available from: \url{http://urn.kb.se/resolve?urn=urn:nbn:se:liu:diva-166431}

\bibitem{SigProc} F. Gustafsson and F. Gunnarsson, ``Mobile positioning using wireless networks: possibilities and fundamental limitations based on available wireless network measurements,'' IEEE Signal Process. Mag., vol. 22, no. 4, pp. 41–53, July 2005.

\bibitem{sensorfusion} F. Gustafsson. Statistical Sensor Fusion. Studentlitteratur AB, third edition, 2018. ISBN 9789144127248.

\bibitem{gdop} D. J. Torrieri. Statistical theory of passive location systems. IEEE Transactions on Aerospace and Electronic Systems, AES-20(2):183–198, 1984.

\bibitem{lowest_gdop} N. Levanon. Lowest GDOP in 2-D scenarios. Radar, Sonar and Navigation, IEEE Proceedings, 147:149 – 155, 07 2000.

\end{thebibliography}
\end{document}